\documentclass[%
 preprint, 
superscriptaddress, aps, pre
 amsmath,amssymb,
 aps, physrev,
]{revtex4-2}

\usepackage{floatrow}
\usepackage{etoolbox}
\usepackage{ulem}

\makeatletter
\patchcmd\captionlabel{\caption@setsubtype*{\FR@tmp}}{\caption@setsubtype*}{}{}
\makeatother

\newfloatcommand{capsidebox}{figure}[\capbeside\useFCwidth][\FBwidth]

\DeclareCaptionLabelFormat{rparen}{\noindent\llap{(\emph{#2})\,}}
\usepackage{adjustbox}

\makeatletter
\newcommand\footnoteref[1]{\protected@xdef\@thefnmark{\ref{#1}}\@footnotemark}
\makeatother

\usepackage{amsthm,amsmath,amssymb}
\usepackage{mathrsfs}

\usepackage{algorithm}
\usepackage{algpseudocode}
\usepackage{amsmath}

\usepackage{amsthm}

\usepackage{mathtools}

\usepackage {graphicx}
\usepackage{caption}
\usepackage{epsfig}
\usepackage{epstopdf}
\usepackage{subcaption}
\usepackage{amsmath,bm}
\usepackage{color}
\usepackage{lmodern}
\usepackage[T1]{fontenc}
\usepackage[utf8]{inputenc}
\usepackage{geometry}
\usepackage{subcaption}
\usepackage{bm}%
\usepackage[colorlinks=true,urlcolor=blue,citecolor=blue,linkcolor=blue]{hyperref}%
\usepackage{tabularx} 
\usepackage{booktabs}

\usepackage{multirow}

\usepackage{threeparttable}

\usepackage{everypage}

\usepackage{makecell}

\AddEverypageHook{\setcounter{footnote}{0}}

\expandafter\ifx\csname package@font\endcsname\relax\else
 \expandafter\expandafter
 \expandafter\usepackage
 \expandafter\expandafter
 \expandafter{\csname package@font\endcsname}%
\fi
\usepackage{verbatim}

\usepackage{xcolor}

\usepackage{float}
\usepackage{subcaption}
\usepackage[percent]{overpic}

\usepackage{booktabs}
\usepackage{siunitx}

\begin{document}


\title{\textbf{Efficient, Geometry-only Prediction of Advection-dominated Particle Transport through Saturated Soils}}

\author{Hao Liu}
\affiliation{Theoretical Physics VI, University of Bayreuth}

\author{Rizwan Khaleel}
\affiliation{Ecosystem Research, Institute of Geography, Faculty of Mathematics and Natural Sciences, University of Cologne}

\author{Markus Rolf}
\affiliation{Ecosystem Research, Institute of Geography, Faculty of Mathematics and Natural Sciences, University of Cologne}

\author{Martin G. J. Löder}
\affiliation{Chair of Animal Ecology I and BayCEER, University of Bayreuth}

\author{Christina Bogner}
\affiliation{Ecosystem Research, Institute of Geography, Faculty of Mathematics and Natural Sciences, University of Cologne}

\author{Stephan Gekle}
\email{stephan.gekle@uni-bayreuth.de}
\affiliation{Theoretical Physics VI, University of Bayreuth}


\begin{abstract}
The environmental transport of microplastic particles has raised concerns because of their potential ecological and human-health impacts.
A particular challenge for experiments on microplastic transport in soils is the fact that soil is a highly complex and non-transparent porous medium. 
Thus, our knowledge of transport mechanisms of microplastics in soils and related ecological implications is still limited and computational simulations offer an additional possibility to obtain important insights into the physical transport behaviour of microplastics.
Yet, direct particle-resolved simulations in realistic soil structures for macroscopic sample sizes remain a computational challenge.
We introduce an efficient pore-network model, minD-PNM, that predicts particle breakthrough curves in water-saturated soils using experimental X-ray micro-computed tomography ($\mu$CT) pore geometries as the only input. 
The model combines an approximately $t^{-3}$ pore-scale transit time distribution with a minimum-dissipation partition of multi-inlet–multi-outlet pores, and propagates these distributions through the pore network to predict breakthrough curves.
We benchmark minD-PNM against lattice Boltzmann and immersed boundary simulations for small systems, and against microplastic column experiments in quartz sediments for macroscopic sample sizes.
In both cases, the predicted breakthrough curves agree well with the corresponding reference data.
Because it requires only geometric information, minD-PNM scales to macroscopic sample volumes offering a practical tool for contaminant-hydrology studies of particulate pollutants.

\textbf{Keywords:} porous media, soil, particle transport, breakthrough curve, pore-network model
\end{abstract}

\maketitle
\section{\label{sec:introduction} Introduction}

Microplastic particles (MPs) are persistent particulate pollutants that have been detected across multiple environmental compartments, including rivers, marine systems, the atmosphere, soils, and groundwater. Once released, they can be redistributed by flowing water, atmospheric transport, sediment resuspension and infiltration, allowing them to migrate away from their original sources and accumulate in environmental reservoirs \citep{Panno2019,Mu2022,Nizzetto2016}. 
MPs are potentially harmful to ecosystems and human health \citep{Koelmans2016,Koelmans2019,WHO2019,Prata2020,li2023potential}. Understanding their transport pathways is therefore essential for assessing their long-term fate and their impact on soils, groundwater and aquatic ecosystems. 

Among these microplastic transport pathways, the soil subsurface is particularly difficult to access by direct observation because soil is optically opaque, so particle trajectories within this matrix cannot be tracked by visual methods. At the same time, the subsurface provides an important transport pathway connecting surface contamination to deeper soil layers and groundwater.
Soils and sediments are porous media whose void space forms a connected but highly irregular network of pores and throats through which water flows. During rainfall, irrigation, or groundwater recharge, suspended MPs are advected by infiltrating water through this structure. Transport through such a medium is characterized at the scale of a bounded, water-saturated sample volume, for which a commonly used observable is the breakthrough curve (BTC), defined as the outlet particle concentration normalized by the inlet concentration as a function of time, measured as the number of injected pore volumes. 
Within the pore space, heterogeneous flow paths, different pore-scale velocity, and slow near-wall regions generate a broad distribution of particle transit times and consequently spread the breakthrough response in time. Particle transport can additionally be affected by retention mechanisms including particle-surface attachment, straining, aggregation and subsequent remobilization \citep{Bradford2002,Shen2008,Bradford2011,Molnar2015}. The present work addresses the transport regime in which particles remain sufficiently well suspended, particle-surface attachment and aggregation are weak, and particle sizes are sufficiently small relative to the accessible pore constrictions that steric retention does not dominate. 
Under these conditions, retention is subordinate to advective transport, and the temporal spreading of micron-sized particles that traverse the sample is governed primarily by the heterogeneous velocity field imposed by the pore geometry.

Current approaches for describing particle transport in water-saturated porous media span different levels of spatial resolution, from high-fidelity direct simulations to reduced pore-network models. 
Direct numerical simulations and particle-resolved pore-scale models based on X-ray micro-computed tomography ($\mu$CT) images can resolve detailed hydrodynamics, particle trajectories, and particle-surface interactions. These approaches provide high physical fidelity, but are typically restricted to relatively small samples because resolving the pore-scale flow field and long particle trajectories in realistic geometries requires substantial memory and computational time \citep{Zhang2019}.
Pore-network models (PNMs) provide a computationally efficient alternative by representing the void space as a network of interconnected pores and throats extracted from $\mu$CT geometries \citep{Blunt2013}. Conventional PNMs efficiently determine network connectivity, pore pressures, and mean throat fluxes, but do not resolve the velocity distribution within individual pores. 
Consequently, they cannot provide the intra-pore transit time distributions of suspended particles which are required for the prediction of breakthrough curves. 
Existing PNM-based transport approaches must therefore introduce additional empirical transit-time closures, perform supplementary pore-scale velocity or streamline calculations or rely on calibration \citep{zhao2023pore,Weishaupt2020,pereira2015time}. 
A predictive relation between pore geometry and intra-pore transit time distributions is therefore needed to obtain breakthrough curves efficiently without resolving the full pore-scale velocity field.

In this work, we develop a minimum-dissipation pore-network model (minD-PNM) for predicting advection-dominated particle transport through water-saturated porous media from $\mu$CT-derived pore geometry. The image-processing and pore-network extraction procedures are described in Section~\ref{sec:method}. The model, introduced in Section~\ref{sec:model}, is based on three main theoretical components. First, for pores with a single inlet and outlet, we derive a flux-weighted transit time distribution and show that, after normalization by the pore-volume transit time, its distribution is well approximated by a $t^{-3}$ scaling over a broad range of pore geometries. Second, for realistic pores with multiple inlets and outlets, we introduce a minimum-dissipation optimization that partitions the pore flux into effective one-in-one-out transport subdomains and assigns a corresponding transit time distribution to each connection. Third, these pore-scale distributions are propagated through the pore network using a flux-weighted random walk to obtain the transit time distribution of the complete sample, which is subsequently convolved with the prescribed inlet concentration history to predict the breakthrough curve. In Section~\ref{sec:results}, we validate the model at two scales:
(i) against particle-resolved lattice-Boltzmann immersed-boundary simulations for small porous samples measured by $\mu$CT and (ii) against microplastic breakthrough experiments in macroscopic quartz-sediment columns. 
By avoiding explicit resolution of the pore-scale velocity field and individual particle trajectories, minD-PNM drastically reduces computational cost and enables transport predictions for considerably larger porous domains.

\section{\label{sec:method} Method}

\subsection{\label{subsec:microct_pnm} $\mu$CT image processing and pore-network construction}

\begin{figure}
  \centering
  \includegraphics[width=0.75\linewidth]{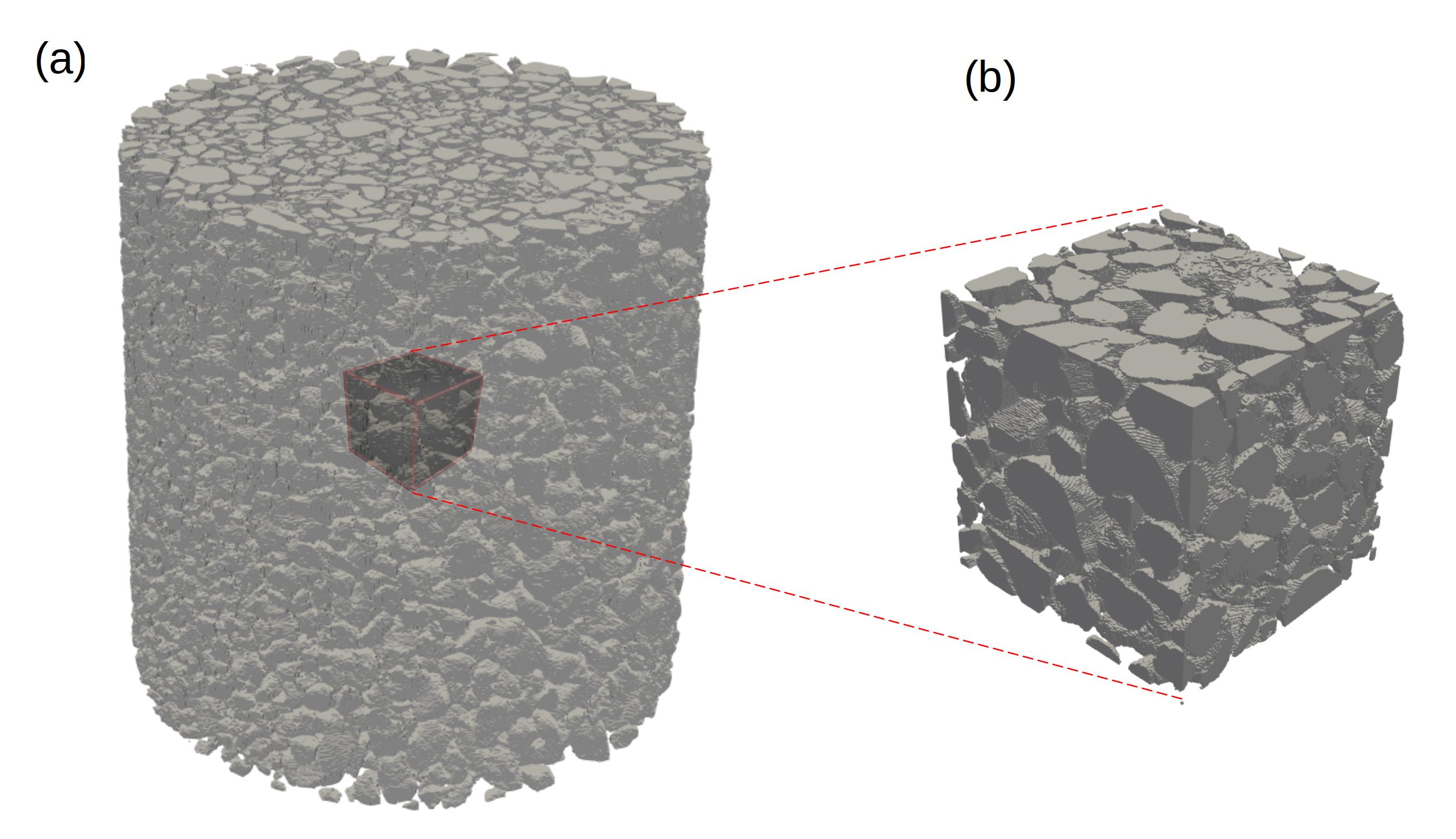}
  \caption{\centering (a)\,Porous medium of quartz sediment column with diameter $880$ voxels and height $866$ voxels from $\mu$CT scanning with porosity $40.5\%$.
    (b) small sample of size $200$ voxels extracted from the center of the column with porosity $35.3\%$ used for full numerical reference simulations.}
  \label{fig:geometryPorousMedia}  
\end{figure}

The porous medium was scanned using a Phoenix v|tome|x s 240. Then its geometry was extracted from the obtained micro-computed tomography ($\mu$CT) images of quartz sediment samples with a resolution of $6.5\,\mu m$. 
The reconstructed grayscale images were converted into solid and pore phases by selecting an appropriate threshold, thereby generating a binary voxel geometry as shown in Fig.~\ref{fig:geometryPorousMedia}.
In addition to the full column geometry shown in Fig.~\ref{fig:geometryPorousMedia}~(a), we consider a small cube of size $200\times6.5\mu m$ extracted from the center of the column for the full numerical reference simulations in \ref{subsec:numerical_validation}.

To convert this voxelized pore space into a pore network, we used a distance-transform-based watershed segmentation \cite{Gostick2017SNOW}. 
The distance transform assigns to each pore voxel its minimum distance to the solid phase.
Local maxima of this distance field define markers for pore centers, from which watershed regions are grown until they meet at watershed boundaries. 
These regions partition the pore space into distinct catchment basins, each corresponding to an individual pore body, while shared boundaries between neighboring basins define throats.
The resulting pore-network model represents the connected void space as a graph, in which pore bodies are nodes and throats are edges. 
For each pore and throat, the network extraction provides geometric and topological quantities, including pore volume, pore connectivity, throat length and throat cross-sectional area \cite{gostick2016openpnm}.

\subsection{Experimental determination of microplastic breakthrough curves}\label{subsec:experimental_validation}

\begin{figure}
  \centering
  \includegraphics[page=1,width=\textwidth]{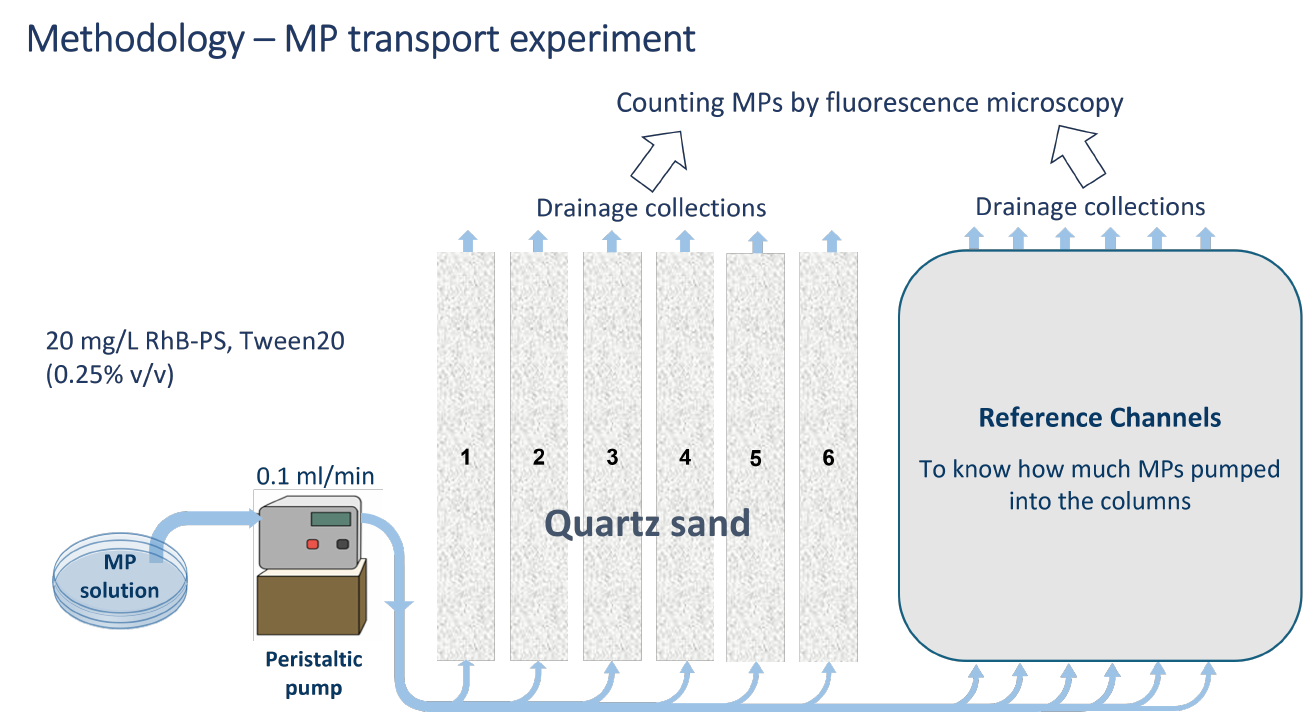}
  \caption{Schematic diagram of the soil column setup used for microplastic transport experiments conducted in the lab to elucidate the breakthrough curves. Blue arrows indicate the flow direction.}
  \label{fig:Methodology_Transport_Experiment}  
\end{figure}

Experimental reference data were obtained from laboratory transport experiments in water-saturated quartz sand columns (Fig.~\ref{fig:Methodology_Transport_Experiment}). Cylindrical columns (12~mm inner diameter, 71.44~mm height; $n=6$) were packed with quartz sand (bulk density~1.7~g\,cm$^{-3}$), fully saturated with artificial rainwater (pH~5.7, \autoref{tab:artificial_rainwater} in Appendix~\ref{app:transport_experiment}), and connected to a calibrated multichannel peristaltic pump (Masterflex LLC, Barrington, IL, USA). A suspension of Rhodamine B-stained polystyrene (RhB-PS) microplastic particles (20~mgL$^{-1}$; particle-size distribution of $D_{90}=10~\mu$m, (see Fig.~\ref{fig:Particle_Size_Distribution} in Appendix~\ref{app:transport_experiment}) in artificial rainwater was used as a stock solution. To minimize the aggregation of RhB-PS particles \citep{eitzen2019} and maintain a homogeneous suspension, 0.25\% (v/v) Tween 20 was added as a dispersant. The suspension was continuously stirred using a magnetic stirrer throughout the experiment to ensure a uniform particle distribution before being injected into the columns under constant flow rate of 0.10 mL\,min$^{-1}$. 

In parallel, blank reference channels not connected to the columns were used to quantify the influent microplastic concentration for each pumping sequence. The experiment consisted of a loading phase, during which the RhB-PS suspension was injected into the columns at a constant flow rate of 0.10 mL\,min$^{-1}$, followed by a flushing phase in which artificial rainwater was pumped through the columns at the same flow rate to elute the remaining microplastic particles. Effluent samples were collected at fixed time intervals during both the loading and flushing phases. The number of microplastic particles in each collected fraction was quantified by fluorescence microscopy using a ZEISS Axio Zoom.V16 microscope (Carl Zeiss Microscopy GmbH, Jena, Germany), and the corresponding particle concentrations were used to construct the experimental breakthrough curves. The breakthrough curves were expressed as the normalized outlet concentration ($C/C_\mathrm{0}$), where $C$ is the measured outlet particle concentration and $C_\mathrm{0}$ is the corresponding influent particle concentration determined from the blank reference channel for each pumping sequence, plotted as a function of injected pore volume. The experimentally determined breakthrough curves were used as independent reference data for model validation.

\subsection{\label{subsec:numerical_validation} Numerical reference data}

\begin{figure}
    \centering
    \includegraphics[width=0.5\linewidth]{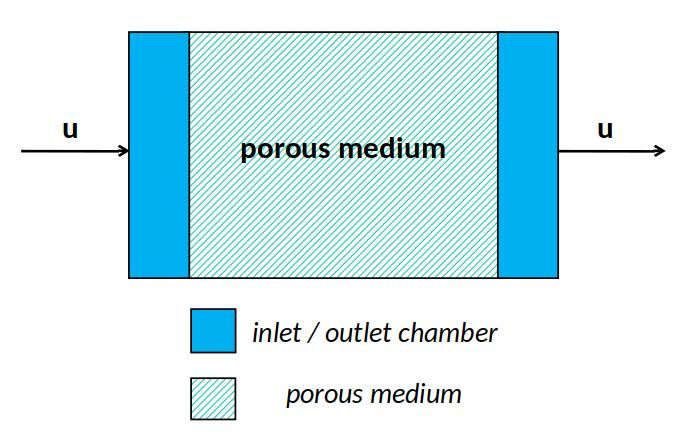}
    \caption{Schematic representation of the 3D simulation domain. The porous medium is bounded upstream and downstream by inlet and outlet chambers, respectively. A prescribed mean inlet velocity $u$ drives the flow through the porous medium and toward the outlet.}
    \label{fig:simulation_geometry}
\end{figure}

For numerical reference data, we use the same binary voxel-based geometric $\mu$CT domain, but consider only a small cube extracted from the center as shown in Fig.~\ref{fig:geometryPorousMedia}~(b). 
The fluid flow was simulated using a lattice Boltzmann method \cite{kruger2017lattice, lehmann2021ejection, Lehmann.20229da}, and the motion of microplastic particles was resolved using an immersed-boundary method. The simulation was executed on the NVIDIA H100 Tensor Core GPU. In the direct numerical simulations, inlet and outlet chambers were artificially added upstream and downstream of the porous sample to impose well-defined flow boundary conditions. 
A prescribed inflow flux was imposed at the inlet chamber, and an equal outflow flux was imposed at the outlet chamber, thereby generating steady saturated flow through the porous medium. 
Periodic boundary conditions were applied in the two transverse directions. 
The simulation geometry is shown in Fig.~\ref{fig:simulation_geometry}.
In practice, particles were injected from the inlet plane using a discrete time protocol. The time step was $\Delta t=0.1\,\mathrm{PV}$, where $\mathrm{PV}$ denotes the injected fluid volume normalized by the pore volume of the porous medium. At each injection time $t_\mathrm{m}=m\Delta t$, with $m=0,1,\ldots,49$, one batch of $N_\mathrm{b}=5000$ particles was released from the inlet plane, giving $50$ injected batches in total. During each time interval $[t_\mathrm{m},t_\mathrm{m+1})$, the number of particles arriving at the outlet was counted. Particles arriving in a given interval may originate from either the current injection batch or from earlier batches. The relative outlet concentration was then calculated by normalizing the number of particles arriving in each interval by the number of particles injected per batch, $N_\mathrm{b}$. The resulting relative outlet concentration as a function of pore volume was used as the breakthrough curve.
These simulations serve only as reference data for validation of our reduced model.

\section{\label{sec:model} Model}

To predict microplastic transport from geometry alone without particle-resolved simulation, we develop an improved PNM approach by introducing a geometry-based approximation for the pore-scale transit time distribution. 
For this, we start from the network geometry extracted from $\mu$-CT images and compute pressures and mean flow rates throughout the network in section~\ref{subsec:pore_pressure_flux}. 
Then, we develop an analytical expression for the generic form of transit time distributions in pores with a single inlet and a single outlet (''one-in-one-out'') in section \ref{subsec:TTD_oneinoneout}.
In section \ref{subsec:subdomain_division} we treat more complex pores with multiple inlets and outlets, for which we estimate local transit times by partitioning complex pores into subdomains and solving a minimum dissipation (minD) optimization problem. 
Then, in section \ref{subsec:random_walk}, these pore-scale transit time distributions are embedded into a random-walk framework that replaces explicit particle tracking and propagates particles through the full network to sample the global transit time distribution. 
Finally, we convert the resulting transit time distribution into a breakthrough curve for the entire medium by temporal convolution in section \ref{subsec:convolution}. 

\subsection{\label{subsec:pore_pressure_flux} Computation of pressures and flow rates in the pore-network model}

Using the pore network constructed in section~\ref{subsec:microct_pnm}, we first compute the graph-scale hydraulic field that supplies the fluxes for the subsequent transit time computations.
Each pore is assigned a single pressure, and each throat carries a steady volumetric flux. 
For a throat connecting pores $i$ and $j$, the flux $Q_{ij}$ is described by
\begin{equation}
    Q_{ij}=g_{ij}(p_i-p_j),
    \label{throat_ij}
\end{equation}
where $p_i$ and $p_j$ are the pore pressures, and $g_{ij}$ is the hydraulic conductance estimated from the throat geometry as follows.
Note that fluxes can be positive or negative and $Q_{ij}=-Q_{ji}$.
Assuming an equivalent cylindrical throat with length $L_{ij}$ and cross-sectional area $A_{ij}$ as obtained from the pore-network construction in section~\ref{subsec:microct_pnm}, the conductance scales as
\begin{equation}
    g_{ij} = \frac{A_{ij}^2}{8\pi \mu L_{ij}}
\end{equation}
with the fluid viscosity $\mu$.

For each internal pore, mass conservation requires the balance of inflows and outflows
\begin{equation}
    \sum_{j} Q_{ij}=0 , \quad \forall i.
\end{equation}
Substitution of the conductance relation gives a sparse linear system for the unknown pore pressures.

To drive the flux, we prescribe an arbitrary, fixed pressure in inlet and outlet pores.
We then solve the pressure problem given above. 
Since the system is linear, the resulting pressure field and throat fluxes are then rescaled so that the total flux matches the prescribed value used in the experimental or full simulation reference data. 
After this rescaling, the pressure drop and all throat fluxes are determined uniquely. 
This graph-level hydraulic solution provides the flux partitioning required for the subdomain division and for the flux-weighted random walk introduced below.

\subsection{Derivation of the transit time distribution in a one-in–one-out pore\label{subsec:TTD_oneinoneout}}

To quantify particle transport through a single pore with one inlet and one outlet, we derive the transit time probability distribution under flux-weighted inlet conditions. Flux-weighted injection means that the probability of releasing a particle on a given area element of the inlet cross section is proportional to the local volumetric flux through that area element. Therefore, faster-flowing regions of the inlet receive more particles than slower-flowing regions. 

\begin{figure}[H]
    \centering
    \begin{subfigure}[b]{0.7\linewidth}
        \centering
        \includegraphics[width=\linewidth]{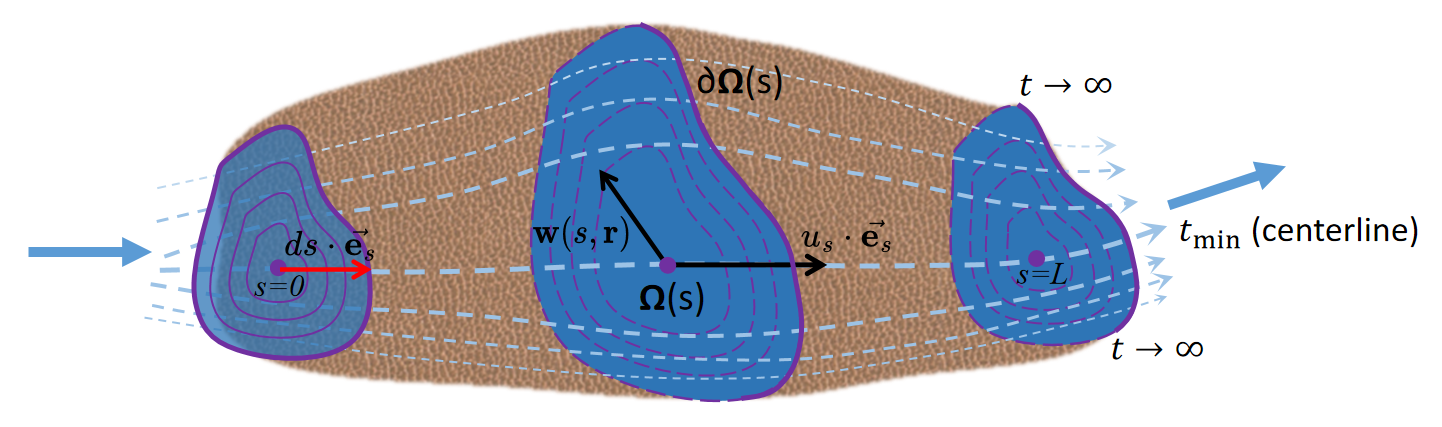}      
    \end{subfigure}
    \caption{\centering Schematic of the Stokes flow through a one-in-one-out pore defining some of the quantities used in the model.}
    \label{fig:sketch_singlePore}
\end{figure} 

We consider steady, incompressible Stokes flow through a one-in–one-out pore whose cross-sectional geometry varies slowly along its axes as illustrated in figure~\ref{fig:sketch_singlePore}. 
The governing equations are:
\[
-\nabla p+\mu\nabla^2\mathbf u=\mathbf 0,\qquad \nabla\!\cdot\!\mathbf u=0.
\]
Let $s$ denote the arc length coordinate along the pore centerline and $\mathbf r$ the transverse coordinate in the cross-section perpendicular to the centerline. 
The velocity is decomposed into an axial component along the centerline and a transverse component normal to it:
\[
\mathbf{u}(s,\mathbf r)=u_s(s,\mathbf r)\,\vec{\mathbf e}_s+\mathbf{w}(s,\mathbf r),
\]
where $\vec{\mathbf e}_s$ is the unit tangent along the centerline and $\mathbf{w}(s,\mathbf r)$ is the transverse velocity in the cross sectional plane at position $\mathbf{r}$.

Under the assumptions of slow cross-section variation and small centerline curvature, we introduce $\varepsilon$ as a small dimensionless parameter defined as the ratio between the characteristic transverse size $R_\mathrm{t}$ and the pore length $L$, i.e. \(\varepsilon \sim \dfrac{R_\mathrm{t}}{L}\). Then, we adopt the scalings of axial velocity and velocity differentials:
\[
u_s \sim U, \,\quad
\partial_s \sim \frac{\varepsilon}{R_\mathrm{t}} ,\quad \partial_{ss}\sim \frac{\varepsilon^2}{R_\mathrm{t}^2} ,\quad
\partial_{\mathbf r}\sim \frac{1}{R_\mathrm{t}},\quad \partial_{\mathbf r\mathbf r}\sim \frac{1}{R_\mathrm{t}^2},
\]
where $U$ is the characteristic axial velocity scale. 
Substituting these scalings into the continuity equation, we get the scale of magnitude of transverse velocity: 
\begin{align}
\partial_s u_s + \nabla_{\mathbf r}\cdot \mathbf w &\sim \frac{\varepsilon U}{R_\mathrm{t}}+\frac{\mathbf{w}}{R_\mathrm{t}} =0\\
\Rightarrow|\mathbf{w}|&\sim \mathcal \varepsilon U.
\end{align}
We then introduce the perturbation expansions:
\begin{align}
    u_s &= u_{0}+\varepsilon u_{1}+\varepsilon^2\,u_2+\mathcal O(\varepsilon^3)\\
\mathbf w &= \varepsilon \mathbf w_1+\varepsilon ^2\,\mathbf w_2+\mathcal O(\varepsilon^3) \\
p&=p_0+\varepsilon p_1+\varepsilon^2 \, p_2+\mathcal O(\varepsilon^3).
\end{align}
Consequently, at order $\mathcal O(1)$, the axial momentum equation is:
\begin{equation}
-\frac{\mathrm d p_0}{\mathrm d s}(s)+\mu\,\nabla_{\mathbf r}^2 u_{0}(s,\mathbf r)=0\; \,\text{in }\;\Omega(s),\quad u_0=0\; \,\text{on }\;\partial\Omega(s)
\label{eq:leadingPoisson}
\end{equation}
where $\Omega(s)$ is the cross-section domain at centerline coordinate $s$ and $\partial\Omega(s)$ is its boundary, \(\nabla_{\mathbf r}^2\) denotes the transverse Laplacian with respect to $\mathbf r$ and $\mathbf{r}$ indicates a point coordinate on $\Omega(s)$.

We define the velocity shape function $\tau(s,\mathbf r)$ and decompose $u_{0}(s,\mathbf r)=G(s)\,\tau(s,\mathbf r)$, where $G(s)$ is the corresponding cross-sectional constant, which is independent of $\mathbf{r}$. Substituting into Eq. (\ref{eq:leadingPoisson}), we obtain a Poisson equation with Dirichlet boundary condition:
\begin{equation}
    -\nabla_{\mathbf r}^2 \tau(s,\mathbf r)=1\ \;\, \text{in }\;\Omega(s),\quad \tau=0\ \;\, \text{on } \;\partial\Omega(s),
    \label{eq:poisson}
\end{equation}
and $G(s)\equiv -\dfrac{1}{\mu}\,\dfrac{\mathrm d p_0}{\mathrm d s}(s)$. We next consider the cross-sectional area $A(s)=|\Omega(s)|$ and define the area averaging operator $\langle \cdot \rangle_{A(s)}=\dfrac{1}{A(s)}\int_{\Omega(s)}\cdot\,\mathrm dA$.
Since the cross-sectional flux
\[
Q(s)=\int_{\Omega(s)} u_{0}\,\mathrm dA
=G(s)\,A(s)\,\langle \tau\rangle_{A(s)}
\]
is constant along $s$, we obtain an alternative form of the cross-sectional constant:
\[
G(s)=\frac{Q(s)}{A(s)\langle \tau\rangle_{A(s)}}.
\]
Then, it is convenient to introduce the normalized velocity profile:
\begin{equation}
\psi(s,\mathbf r):=\frac{\tau(s,\mathbf r)}{\langle \tau\rangle_{A(s)}}, \label{eq:defPsi}
\end{equation}
which satisfies:
\[
u_{0}(s,\mathbf r)=\frac{Q}{A(s)}\,\psi(s,\mathbf r) \quad\text{and} \quad \langle\psi\rangle_{A(s)}=1.
\]

For a uniform particle concentration upstream, the number of particles entering an area element 
$dA\subset\Omega(s)$ during a fixed time interval is proportional to the local normal volumetric flux, 
$u_0(s,\mathbf r)dA$. 
After normalization by the total flux $Q$, the probability density for particle injection on the inlet cross section is therefore $\dfrac{u_0(s,\mathbf r)}{Q}=\dfrac{\psi(s,\mathbf r)}{A(s)}$.
This flux-weighted injection is used throughout the following derivation.

To compare particle transport across pores with different geometries, we introduce a dimensionless transit time.
A natural time scale for transport is the pore volume time
\begin{equation}
t_{\mathrm{PV}} := \frac{V_\mathrm{p}}{Q_\mathrm{p}},
\end{equation}
where $V_\mathrm{p}=\int_0^L A(s)\,\mathrm ds$ is the volume and $Q_\mathrm{p}$ the flow rate. In the following, transit times are expressed in dimensionless pore volume units by
\begin{equation}
t := \frac{t_{\mathrm{SI}}}{t_{\mathrm{PV}}},
\end{equation}
where $t_{\mathrm{SI}}$ denotes the physical transit time measured in SI units.

\subsubsection{Transit time distribution of elliptic Poiseuille flow}

We now analyse how the cross-sectional geometry influences the transit time distribution by considering a Poiseuille-type flow with an elliptic cross section. Since the cross section is constant in the one-dimensional setting, we drop the axial coordinate \(s\) and retain only the transverse coordinate \(\mathbf r\). 

In a straight tunnel of length \(L\) with a fully developed flow, i.e. with velocity profile \(\psi\) independent of \(s\), the transit time in pore-volume units for a particle injected at cross-sectional position \(\mathbf r\) is
\[
t(\mathbf{r}) = \frac{t_{\mathrm{SI}}}{t_{\mathrm{PV}}}
= \frac{L/(G\,\tau(\mathbf{r}))}{L\,A\;/\;(G\int_{\Omega}\tau(\mathbf{r})\,dA)}
=\frac{\langle \tau\rangle_{A}}{\tau(\mathbf r)}
= \frac{1}{\psi(\mathbf{r})}.
\]
We solve the Poisson Eq.~(\ref{eq:poisson}) to obtain $\tau$ and normalize the solution via Eq.~(\ref{eq:defPsi}) to obtain the velocity field \(\psi\), and then compute the transit time distribution from 
\begin{equation}
\eta(t)=\frac{1}{A}\int_{\Omega}\psi(\mathbf r)\,\delta\!\left(t-\frac{1}{\psi(\mathbf r)}\right)\,dA
=
t^{-3}\,\frac{1}{A}\int_{\{\psi=1/t\}} \frac{d\ell}{\left|\nabla\psi(\mathbf r)\right|},
\quad t\ge t_{\min}=\frac{1}{\psi_{\max}},
\label{eq:eta_from_psi}
\end{equation}
where \(d\ell\) is the arc-length element along the corresponding level set. The derivation is provided in Appendix~\ref{app:velocityToTTD}. 

As a reference case, we consider a straight pore with a fully developed Poiseuille profile in a circular cross section. With the velocity field \(\psi(r)=2(1-r^2/R^2)\), the transit time distribution reads
\begin{equation}
\eta(t)=\frac{1}{2}\,t^{-3},\quad t\ge\frac{1}{2}. 
\label{eq:eta_ellipse}
\end{equation}
Extending the circular case to an ideal elliptic cross section yields the same transit time distribution. The derivation is provided in Appendix~\ref{app:ellipticPoiseuille}. 

\subsubsection{Transit time distribution of arbitrary shaped Poiseuille-type flow pore}

Although the $t^{-3}$ law in Eq.~(\ref{eq:eta_ellipse}) is exact only for elliptic pores, it is approximately valid also for more general cross sections.
This can be understood from the local structure of the velocity field, in particular near the maximum-velocity point and near the boundary, where the integral $\frac{1}{A}\int_{\{\psi=1/t\}} \frac{d\ell}{\left|\nabla\psi(\mathbf r)\right|}$ may exhibit the largest deviation.

Near the cross-sectional center, where \(\psi\) reaches its maximum, we have \(\nabla\psi=0\), and a second order Taylor expansion gives
\begin{equation}
    \psi(\mathbf r)\approx \psi_{\max}+\frac{1}{2}(\mathbf r-\mathbf r_\mathrm{c})^{\!\top} H\,(\mathbf r-\mathbf r_\mathrm{c}),
\label{eq:elleptic_velocity_approximation}
\end{equation}
where \(H=\begin{pmatrix}
\partial_{xx}\psi & \partial_{xy}\psi\\
\partial_{yx}\psi & \partial_{yy}\psi
\end{pmatrix}\) is the Hessian matrix at the isolated maximum velocity point $\mathbf r_\mathrm{c}$. Since \(H\) is negative definite, the level sets of \(\psi\) in a neighborhood of \(\mathbf r_c\) are approximately ellipses. Thus, the short-time part of the transit time distribution is locally governed by an elliptic geometry.

On the other hand, near a smooth no-slip boundary, the axial velocity vanishes at the wall and increases approximately linearly with the inward normal distance $y^+$, such that $|\nabla\psi|$ is constant to leading order with corrections of $O(y^+)$. The low-velocity level sets are therefore asymptotically close to the boundary, and their lengths satisfy $\mathcal{L}(y^+)=\mathcal{L}_0+O(y^+),$ where $\mathcal{L}_0$ is the boundary length. Since $t\sim\psi^{-1}\sim(y^+)^{-1}$ and the flux weighting is proportional to $\psi$, the transit-time density scales as  \[\psi \mathcal{L}(y^+)\left|\frac{dy^+}{dt}\right|\sim t^{-1}\left[\mathcal{L}_0+O(t^{-1})\right]t^{-2}=\mathcal{L}_0t^{-3}+O(t^{-4}).\] Thus, the geometric variation of the near-wall level sets contributes only higher-order corrections and does not alter the leading $t^{-3}$ tail. 
However, this argument assumes that, within the asymptotic near-wall region, the low-velocity level sets can be approximated as parallel offsets of the boundary with an approximately uniform normal separation. This approximation may become less accurate where the local wall geometry gradient varies strongly, for example near sharp corners, regions of rapidly varying curvature or cross-sections with sufficiently large aspect ratios. Such geometries can distort the low-velocity level sets away from simple parallel offsets and therefore require separate numerical validation.

To test the above cases, we solve Eq.~(\ref{eq:poisson}) numerically by a finite difference method, normalize the discrete solution to obtain \(\psi\), and then evaluate the resulting transit time distribution from Eq.~(\ref{eq:eta_from_psi}). 
The numerical evaluation becomes less reliable at very large transit times. 
In this regime, the relevant level sets lie very close to the boundary, where the corresponding transit times approach infinity and discretization errors are amplified, making the result increasingly sensitive to the near-wall resolution. 
In the following, we therefore neglect the contribution from a thin near-wall region and focus on the range in which the numerical transit time distribution is sufficiently smooth and robust. 
Details of the numerical method are provided in Appendix~\ref{app:Numerical_method_PDF_singlePore}.

\begin{figure}[H]
    \centering
    \vspace{0.3em}
    \begin{subfigure}[t]{0.45\linewidth}
        \centering
        \begin{overpic}[width=\linewidth]
            {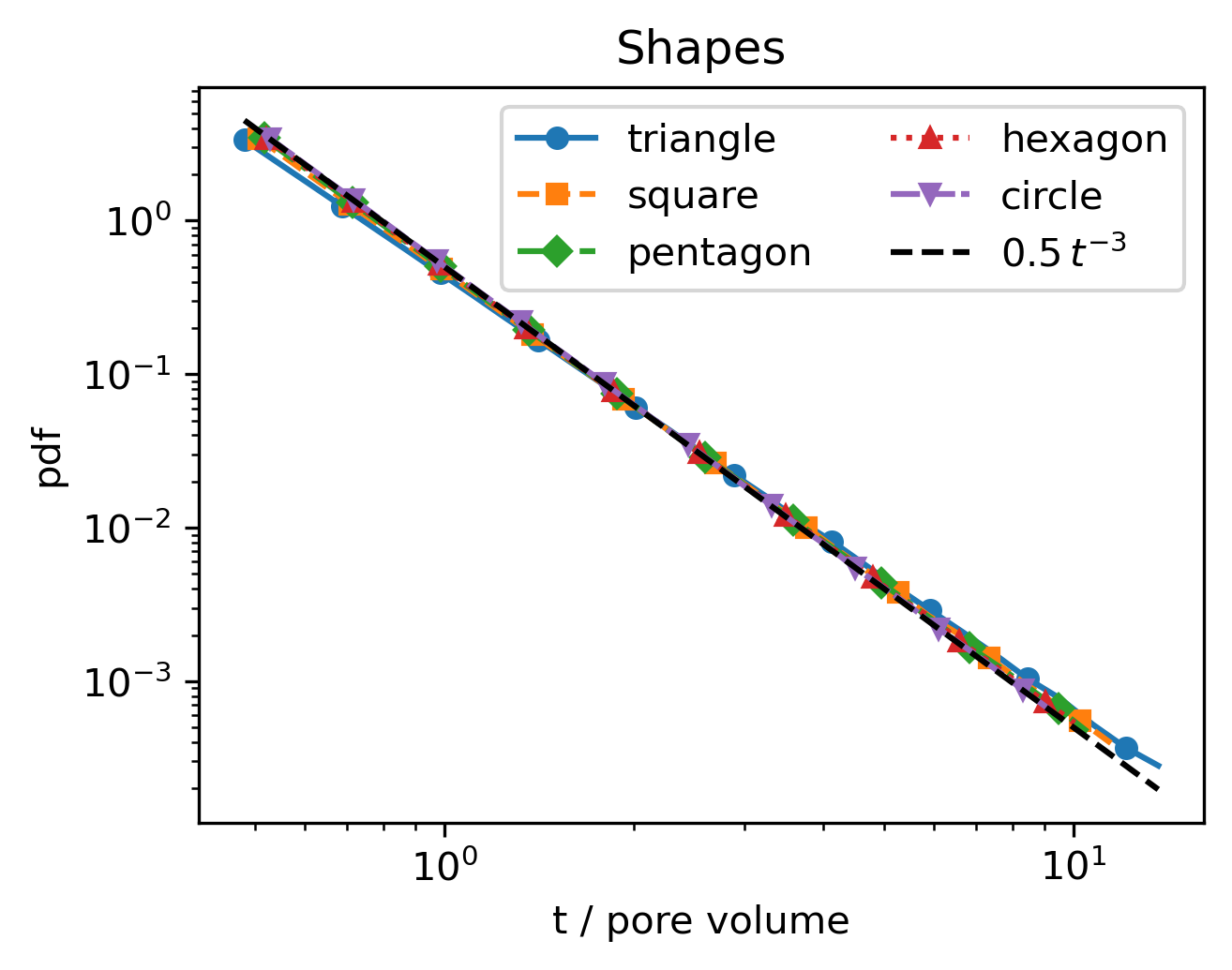}
        \end{overpic}
        \caption{}
        \label{fig:TTD_shapes}
    \end{subfigure}
    \hfill
    \begin{subfigure}[t]{0.45\linewidth}
        \centering
        \begin{overpic}[width=\linewidth]
            {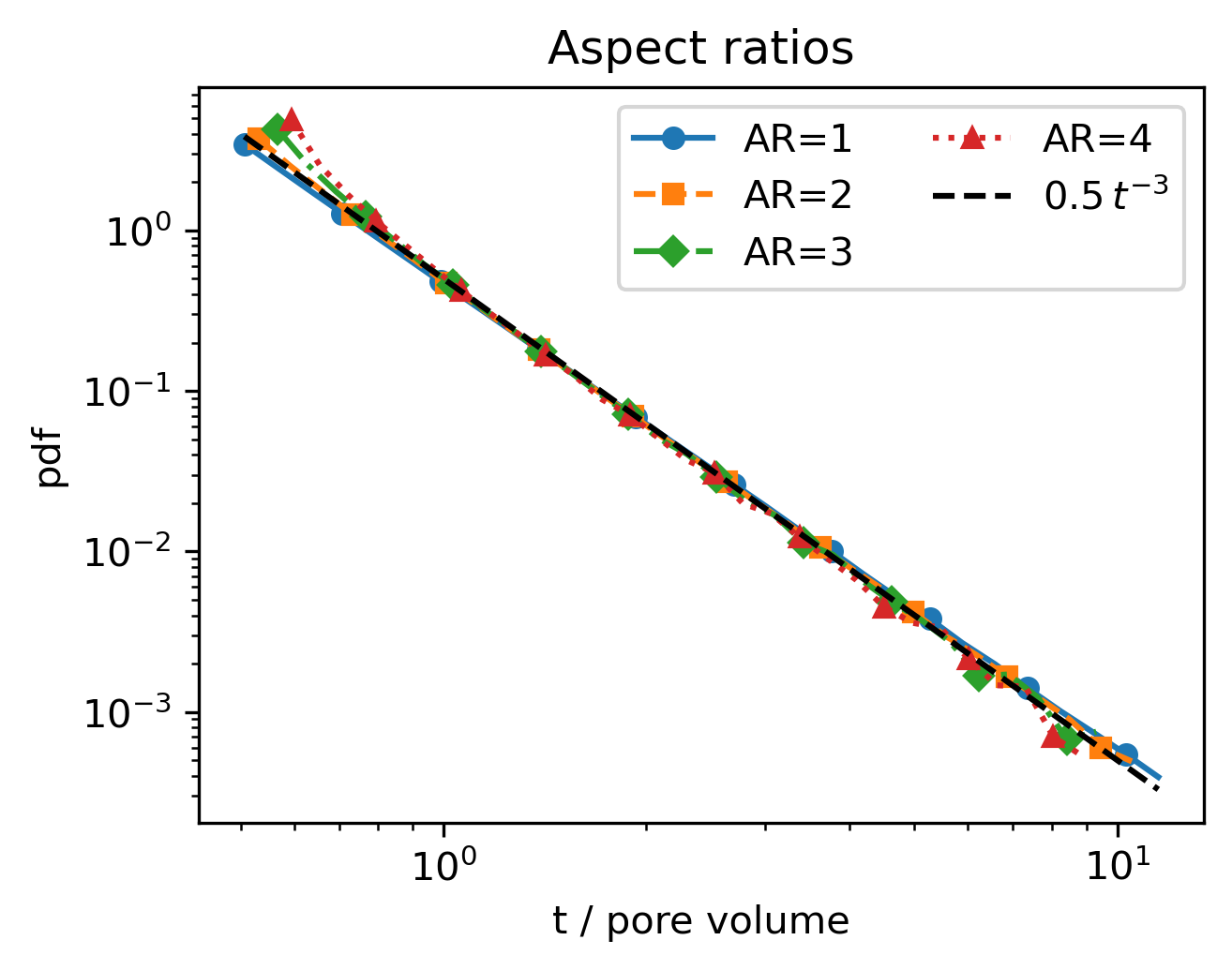}
        \end{overpic}
        \caption{}
        \label{fig:TTD_AR}
    \end{subfigure}
    \vspace{0.1em}
    \begin{subfigure}[b]{0.6\linewidth}
        \centering
        \begin{overpic}[width=\linewidth]
            {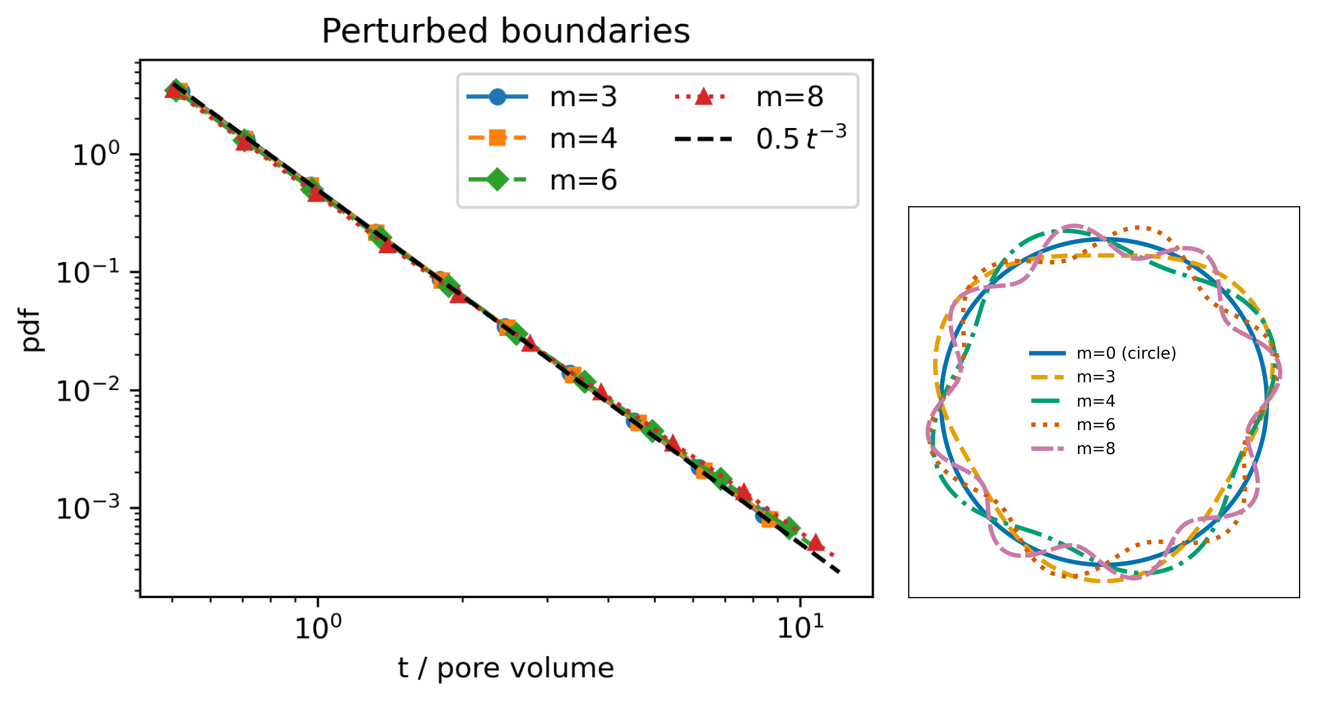}
        \end{overpic}
        \caption{}
        \label{fig:TTD_perturbation}
    \end{subfigure}
\caption{\centering Numerical transit time distributions for one-dimensional Poiseuille flow through pores with complex cross-sectional geometries remain close to the exact analytical result $\dfrac{1}{2}t^{-3}$ including
    (a) regular polygonal shapes; (b) rectangular cross sections with different aspect ratios; (c) sinusoidally perturbed circular boundaries, \(r(\theta)=R\,+\gamma\sin(m\theta)\), with perturbation amplitude \(\gamma=0.1R\).}
    \label{fig:TTD_intermedia_region}
\end{figure}

Fig.~\ref{fig:TTD_intermedia_region} shows numerical transit time distributions for different cross-sectional geometries and assesses the robustness of the \(t^{-3}\) approximation. 
As shown in Fig.~\ref{fig:TTD_shapes}, for regular polygonal cross sections, including triangle, square, pentagon, hexagon, and circle, the numerical transit time distributions remain very close to the approximate reference law. For polygonal cases, the distributions approach the circular limit as the number of edges increases, which confirms the robustness of the \(t^{-3}\) approximation for weakly angular cross sections.
In Fig.~\ref{fig:TTD_AR} we show rectangular geometries with different aspect ratios (AR).
As aspect ratio increases, the most visible deviations appear near the smallest transit times, whereas the larger time ranges, especially the intermediate time ranges, still remain close to the approximated trend. For \(\mathrm{AR}\lesssim 2\), the approximation is very good over nearly the entire resolved range. For larger aspect ratios, the deviations become more noticeable. 
Hence, for moderate aspect ratios, approximately \(\mathrm{AR}\lesssim 4\), the overall agreement remains reasonable, and the \(t^{-3}\) approximation is still robust. 
In Fig.~\ref{fig:TTD_perturbation}, we investigate sinusoidally perturbed circular boundaries.
For perturbation modes \(m=3,4,6,\) and \(8\), the numerical transit time distributions remain in very good agreement with the reference law, indicating that moderate smooth perturbations have only a limited effect on the overall transit time statistics, as long as they do not substantially alter the underlying large scale shape of the cross section.
As a comparison, the wavy boundary with mode \(m=3\) can be viewed as a triangle-like cross section in which the sharp corners have been smoothed out. Its distribution is much closer to the reference curve than that of the triangle, further indicating that the deviation is mainly caused by the presence of sharp corners.

In summary, to simplify the prediction of transit time distributions, we will in the following adopt the approximation
\begin{equation}
    \eta(t)\approx\frac{1}{2}\,t^{-3},\quad t\ge\frac{1}{2},
\end{equation}
which is the same as for an elliptic cross section given in Eq.~(\ref{eq:eta_ellipse}).
This approximation is expected to be accurate for pores without sharp corners, with smooth or moderately perturbed boundaries, and with moderate aspect ratios.

\subsubsection{Transit time distribution in varying pores}
\label{subsec:varyingOneInOneOut}

We now extend the straight-pore result to a general one-in-one-out pore whose cross section varies slowly along the axial coordinate $s$ and whose centreline is weakly curved. 
In such a pore, the exact dimensionless transit time is a Lagrangian quantity defined along a streamline which can be simplified using the following two approximations.

The first approximation is to replace the exact Stokes velocity along each streamline by the local Poiseuille flow associated with the same cross-sectional profile. Under the assumptions of slow axial variation and weak curvature, this local Poiseuille approximation is consistent since axial variation of the cross section and centreline curvature generate only higher-order corrections to the streamline transit time. Therefore, the local flux-weighted transit time statistics retain the same leading $t^{-3}$ structure.

A second approximation is that we compute the transit time by an Eulerian axial accumulation of the local cross-sectional contributions whereas the exact averaging is Lagrangian under streamline-based description. The corresponding error comes from the fact that a fixed Eulerian velocity level does not exactly follow the same streamline when the cross section changes along the pore. This error remains small in a slowly varying pore, because both the axial change of the geometry and the induced fluctuation of the local transit time statistics are weak.

As a result, neither approximation changes the leading power-law form of the transit time distribution. We therefore approximate the transit time distribution in a general one-in-one-out pore by
\begin{equation}
\eta_{\mathrm{ref}}(t) = \frac12\,t^{-3},\quad t\ge \frac12
\label{eq:singlePoreConstantApproximation}
\end{equation}
which we will use as a reference distribution from now on.
 The detailed derivation and the corresponding error estimates are given in Appendix~\ref{App:ErrAnalysisSs}.

\subsection{Subdomain division by minimum dissipation
\label{subsec:subdomain_division}}

In realistic sediments or soils, a single pore often contains multiple inlets and outlets. 
Therefore, an appropriate partitioning strategy is required to divide the pore into subdomains, each containing only one inlet and one outlet, while ensuring flux conservation.

\begin{figure}[H]
    \centering
    \begin{subfigure}[b]{0.5\linewidth}
        \centering
        \includegraphics[width=\linewidth]{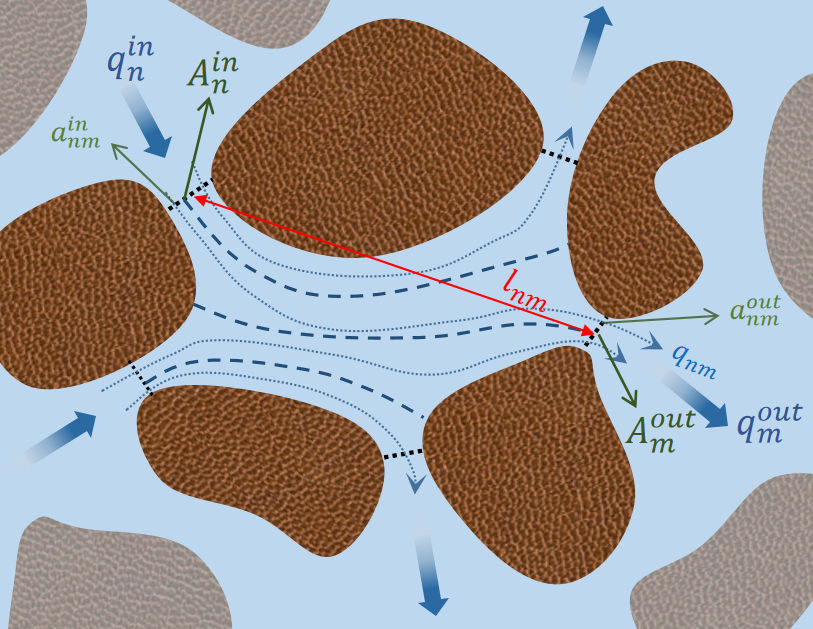}      
    \end{subfigure}
    \caption{\centering Schematic of complex flow in a multi-in-multi-out pore and the subdomain division.}
    \label{fig:sketch_subdomain}
\end{figure} 

We first recall the dissipation principle for pressure-driven laminar flow, which motivates our subdomain division strategy. 
For steady Poiseuille flow in a rigid pipe the volumetric flow rate $q$ is proportional to the pressure drop $\Delta p$, so that $\Delta p = R\,q$,
where \(R\) is the hydraulic resistance. 
Moreover, because the flow is steady, there is no accumulation of kinetic energy within the pipe. 
Hence, the pressure work supplied by the pressure drop is entirely converted into viscous dissipation, and the dissipation rate satisfies
\[
P=\Delta p\,q=R\,q^2.
\]

Now we consider the $i$-th pore with inflows and outflows to all neighbouring pores labelled by $j$ given by $Q_{ij}$ as defined in eq.~(\ref{throat_ij}).
We now introduce a pore-level indexing where $n=1\ldots N$ labels all inflow throats ($Q_{ij}<0$) and $m=1\ldots M$ labels all outflow throats ($Q_{ij}>0$).
Inside the $i$-th pore, we assume that each inlet is connected by a flow channel to each outlet and the corresponding flux is denoted as $q_{nm}^{(i)}$.
In the following, we will indicate the upper index $(i)$ only where it is explicitly needed and omit it otherwise. The total influx through the $n$-th throat $q_n^{\mathrm{in}}$ and the total outflux through the $m$-th throat $q_m^{\mathrm{out}}$ are prescribed by the corresponding network-scale throat fluxes. They are used to constrain the subdomain optimization below.
Geometrically, we define the areas of the inlet throats as $A^\mathrm{in}_n$ and those of the outlet throats as $A^\mathrm{out}_m$. We further introduce an effective length of the subdomain flux channels as $l_{nm}$, which is defined as the Euclidean distance between the centroids of inlet $n$ and outlet $m$.
Then, quantities $q_n^{\mathrm{in}}$, $q_m^{\mathrm{out}}$, 
$A_n^{\mathrm{in}}$, $A_m^{\mathrm{out}}$, and $l_{nm}$ are determined from the pore-network construction described in Section~\ref{subsec:microct_pnm} and the corresponding network-scale hydraulic  calculation described in Section~\ref{subsec:pore_pressure_flux}, and they are treated as known parameters in the following subdomain-flux optimization.
 
Following from the Hagen–Poiseuille law, the resistance in general scales as
$R \sim \mu\, \dfrac{l}{r^4} \sim \mu\, \dfrac{l}{A^2}$.
To assign a resistance cost to the subdomain connecting inlet \(n\) to outlet \(m\), we approximate it as a Poiseuille conduit of length \(l_{nm}\) and effective squared cross-sectional area $A_n^\mathrm{in}A_m^\mathrm{out}$.
We therefore approximate the resistance as:
\begin{equation}
R_{nm}=C_\mathrm{v}\,\frac{l_{nm}}{A_n^{\mathrm{in}}A_m^{\mathrm{out}}},
\label{eq:cost_definition}
\end{equation}
where $C_\mathrm{v}$ is a constant including dynamic viscosity $\mu$. This resistance serves as the energetic cost for flow from inlet $n$ to outlet $m$.

Therefore, the question of subdomain division can be reformulated as an optimization transport problem by minimizing energy dissipation. 
Let the flux allocation matrix be $\mathbf{q} \in \mathbb{R}^{N \times M}$, where each element $q_{nm}$ represents the sub-flux transported from inlet $n$ to outlet $m$. 
The objective is to minimize the total transport cost:
\begin{equation}
\min_{\mathbf{q}} \sum_{n=1}^{N} \sum_{m=1}^{M} R_{nm} q_{nm}^2 ,
\label{eq:ot_objective}
\end{equation}
subject to the flux conservation constraints
\begin{equation}
\sum_{m=1}^{M} q_{nm} = q^{\text{in}}_n, \quad
\sum_{n=1}^{N} q_{nm} = q^{\text{out}}_m, \quad q_{nm} \ge 0.
\label{eq:ot_constraints}
\end{equation}

This convex quadratic minimization problem can be solved using the alternating direction method of multipliers (ADMM) \cite{wahlberg2012admm,boyd2011distributed}.  
Now consider a one-in-one-out subdomain \(n \to m\). The exact cross-sectional areas of the individual flow tubes connecting inlet $n$ to outlet $m$ cannot be determined from the pore geometry. We therefore approximate them using flux-weighted effective areas:
\begin{equation}
  a^{\mathrm{in}}_{nm} \;=\; \frac{q_{nm}}{q^{\mathrm{in}}_{n}}\,A^{\mathrm{in}}_{n},
  \qquad
  a^{\mathrm{out}}_{nm} \;=\; \frac{q_{nm}}{q^{\mathrm{out}}_{m}}\,A^{\mathrm{out}}_{m}.
\end{equation}
Using these effective areas, we approximate the volume of the connection \(n\to m\) as a truncated cone $\widehat V_{nm}=\dfrac{l_{nm}}{3}\,\large(a_{nm}^{\mathrm{in}}+a_{nm}^{\mathrm{out}}+\sqrt{a_{nm}^{\mathrm{in}}a_{nm}^{\mathrm{out}}}\large)$. 
By renormalizing these measures, the effective subdomain volume is defined as
\begin{equation}
V_{nm}=V_\mathrm{p}\,
\frac{\widehat V_{nm}}
{\sum\limits_{n'=1}^{N}\sum\limits_{m'=1}^{M}\widehat V_{n'm'}},
\label{eq:Vij_def}
\end{equation}
which guarantees that $\displaystyle \sum\limits_{n=1}^{N}\sum\limits_{m=1}^{M} V_{nm} = V_\mathrm{p}$ with the pore volume \(V_\mathrm{p}\).
We then define the subdomain volume weight, the subdomain flux weight and their ratio, respectively:
\begin{equation}
  (\omega_V)_{nm} \;=\; \frac{V_{nm}}{V_{\mathrm{p}}},
  \quad
  (\omega_q)_{nm} \;=\; \frac{q_{nm}}{Q_{\mathrm{p}}},
  \quad
  \lambda_{nm}=\frac{(\omega_V)_{nm}}{(\omega_q)_{nm}}=\frac{V_{nm}/q_{nm}}{V_\mathrm{p}/Q_\mathrm{p}},
\end{equation}
where \(Q_\mathrm{p}=\displaystyle\sum_{n=1}^{N} q^{\text{in}}_n\) is the total flux through the pore and \(\lambda_{nm}\) measures the characteristic transit time of the subdomain relative to the mean pore transit time. 
Rescaling the approximated reference distribution~\eqref{eq:singlePoreConstantApproximation}
from the subdomain transit time \(t_\mathrm{sb}\) to the pore-scale transit time
\(t=\lambda_{nm}t_\mathrm{sb}\) requires conservation of probability. 
Specifically, the probability associated with corresponding time intervals must remain
unchanged,
\(\eta_{nm}(t)\,dt
=\eta_{\mathrm{ref}}(t_{sb})\,dt_{sb}\). 
Therefore, the contribution of subdomain \(n\to m\) is given by
\begin{equation}
\eta_{nm}(t)=
\frac{1}{\lambda_{nm}}\,
\eta_{\mathrm{ref}}
\left(\frac{t}{\lambda_{nm}}\right)=
\frac{1}{2}\lambda_{nm}^{2}t^{-3},
\quad
t\geq \frac{\lambda_{nm}}{2}.
\label{eq:etaij_def}
\end{equation}
Using this equation, breakthrough curves can be computed as detailed in the two following sections.

For illustration purposes (see Fig.~\ref{fig:Quartz_Porous_Distribution} below), we  compute a pore-scale transit time distribution by weighting each subdomain distribution by the corresponding flux fraction \((\omega_q)_{nm}\) and summing over all connections yields,
\begin{equation}
\eta_{\mathrm{pore}}(t)=
\sum_{n=1}^{N}\sum_{m=1}^{M}
(\omega_q)_{nm}\,\eta_{nm}(t)=
\frac12\sum_{n=1}^{N}\sum_{m=1}^{M}\frac{(\omega_V)_{nm}^2}{(\omega_q)_{nm}}\,t^{-3}\,
\mathbf{1}_{\left\{t\ge \frac12\,\frac{(\omega_V)_{nm}}{(\omega_q)_{nm}}\right\}},
\label{eq:pdf_pore}
\end{equation}
where \(\mathbf{1}_{\{\cdot\}}\) denotes the indicator function. 

\subsection{Random-walk evaluation of transit times}\label{subsec:random_walk}

Particle trajectories through the pore network are generated using a flux-weighted random walk that retains the identity of the inlet and outlet when particles pass through each pore.
Let $(i)$ denote the $i$-th pore visited along a trajectory.
The outlet is selected randomly according to the conditional transition probability:
\begin{equation}
    P^{(i)}\left(m\mid n\right)=
    \frac{q_{nm}^{(i)}}
    {\displaystyle\sum_{m'=1}^M q_{nm'}^{(i)}}.
    \label{eq:conditional_transition_probability}
\end{equation}
Once the outlet has been selected, the dimensionless transit time in pore $i$ is sampled directly from the corresponding subdomain distribution,
\begin{equation}
    \theta_i
    \sim
    \eta_{nm}^{(i)}(t),
    \label{eq:subdomain_time_sampling}
\end{equation}
where $\eta_{nm}^{(i)}$ is given by Eq.~\eqref{eq:etaij_def}. 
The selected outlet subsequently becomes the inlet of the downstream pore, and the procedure is repeated until the particle reaches the network outlet. At the upstream boundary, particles are injected into the inlet pores in proportion to their imposed inlet fluxes.

For a trajectory that visits $I_N$ pores, the total transit time normalized by the medium-scale pore volume is
\begin{equation}
    T=
    \frac{
        \displaystyle
        \sum_{i=1}^{I_N}
        \frac{V^{(i)}_\mathrm{p}}{Q^{(i)}_\mathrm{p}}\,\theta_i
    }{\displaystyle
       \frac{V_{\mathrm{PM}}}{Q_{\mathrm{PM}}}
    },
    \label{eq:path_transit_time}
\end{equation}
where $V_{\mathrm{PM}}$ and $Q_{\mathrm{PM}}$ are the total void volume and imposed volumetric flow rate of the porous medium, respectively.

\subsection{\label{subsec:convolution} Convolution with the Transit Time Distribution}

To connect the microscopic transit time distributions to the macroscopic breakthrough response, we consider the spatially averaged transport of any given particle-laden flow through the porous medium. 
In this context, both the inlet and outlet represent the total mass flux of particles entering or leaving the domain per unit time.  

The outlet concentration at time $t$ can therefore be expressed as the convolution between the inlet signal and the transit time distribution:
\begin{equation}
C_{\text{out}}(t)
\;=\;
\int_{0}^{\infty} C_{\text{in}}(t-\tau)\,\eta(\tau)\,d\tau
\;=\;
(C_{\text{in}} * \eta)(t),
\label{eq:convolution}
\end{equation}
where $\eta(\tau)$ denotes the probability density function of transit times across all flow paths in the porous medium, and $C_{\text{in}}(t)$ represents the inlet injection concentration. 

Physically, this formulation captures that the outlet response $C_{\text{out}}(t)$ results from the superposition of particles entering at earlier times, each delayed according to the stochastic travel-time distribution $\eta$. 
The convolution thus provides a natural bridge between pore-scale transport dynamics and the macroscopic, spatially averaged breakthrough behavior.

 \section{Results and Analysis \label{sec:results}}

\subsection{Predicting transit time distributions for individual pores}

\begin{figure}[H]
    \centering
    \begin{subfigure}[b]{0.51\linewidth}
        \centering
        \includegraphics[width=\linewidth]{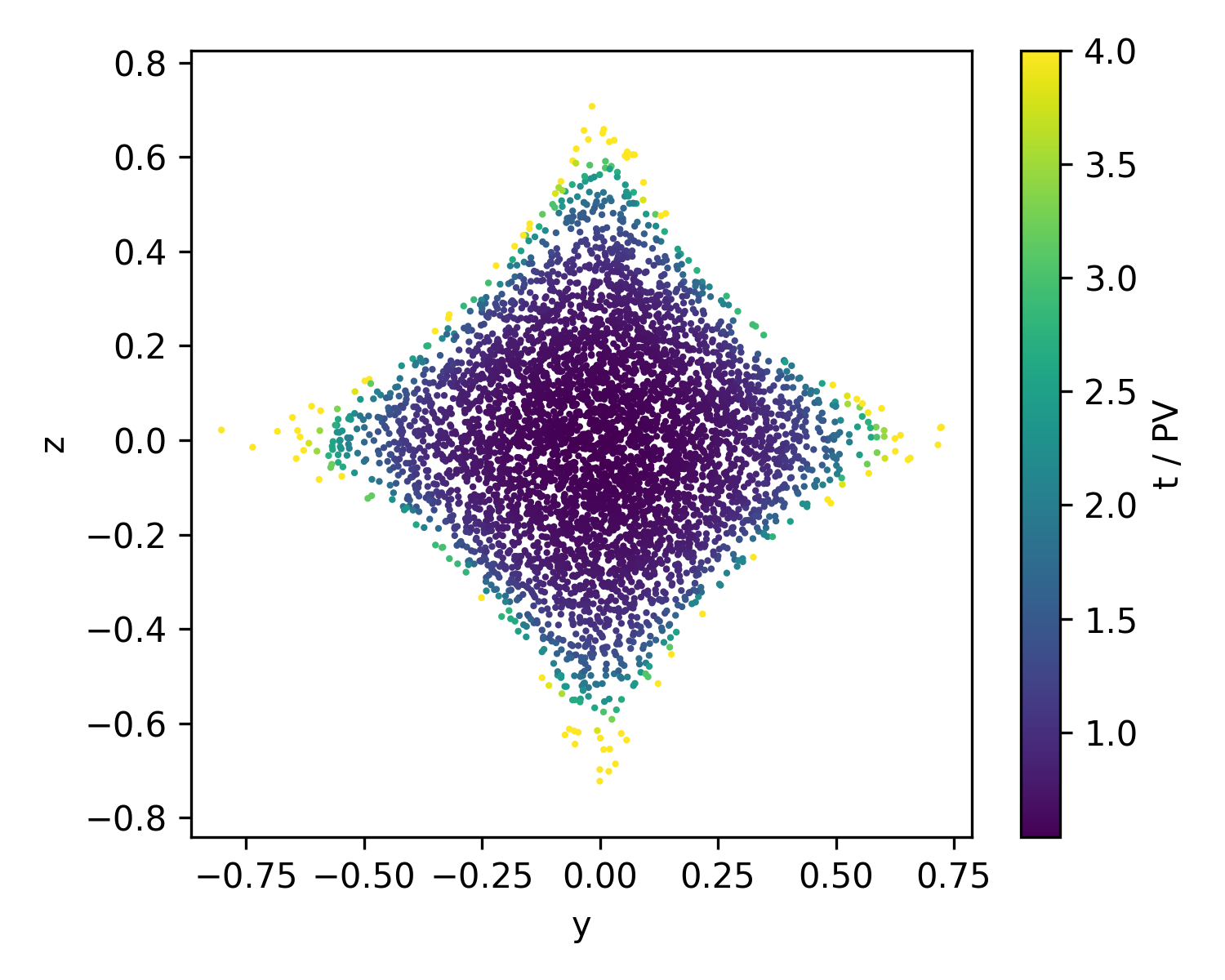}  \caption{}       \label{fig:pv_scatter_sc}
    \end{subfigure}
    \hfill
    \begin{subfigure}[b]{0.48\linewidth}
        \centering
        \includegraphics[width=\linewidth]{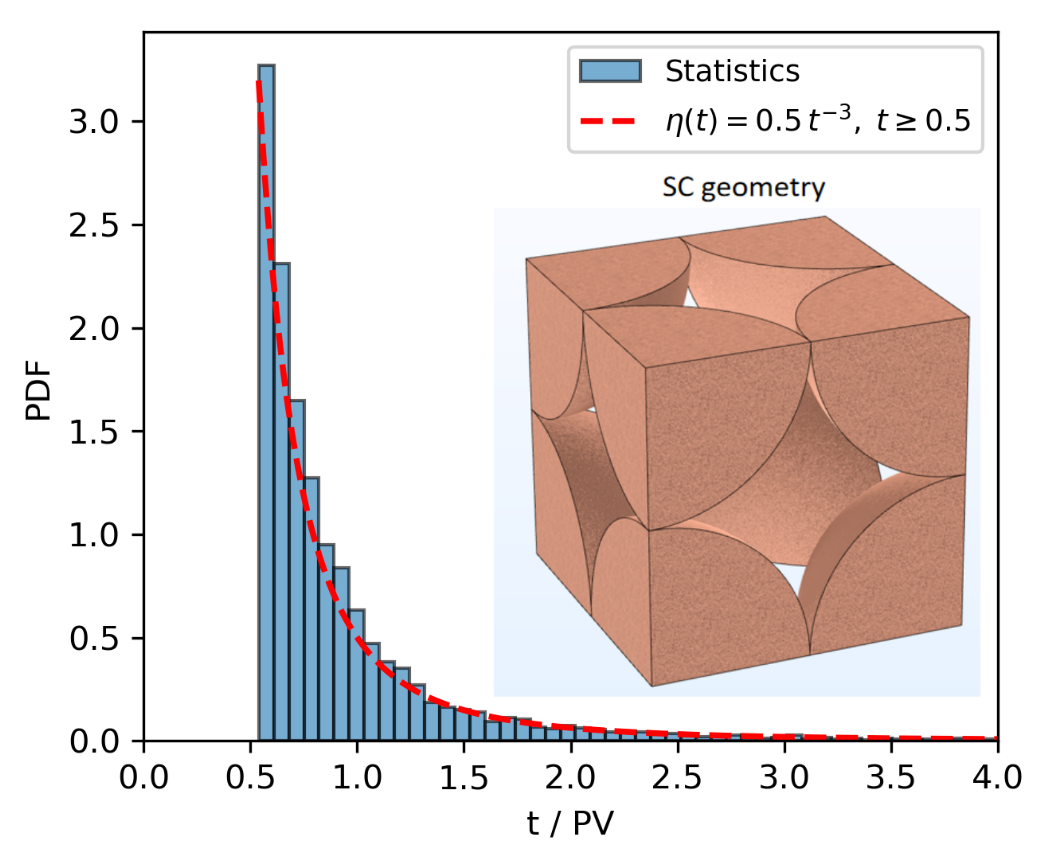}
        \caption{}       \label{fig:pv_pdf_compare_sc}
    \end{subfigure}
    \caption{\centering Transit time distribution of $5000$ particles through an array of spheres on an SC grid: (a) transit time as a function of the flux-weighted inlet position; (b) Statistical distribution and $t^{-3}$-power approximation.}
    \label{fig:SC_statistics_prediction}
\end{figure}

To assess the accuracy of our minD-PNM, we simulate particle transport through a simple cubic (SC) sphere packing using the LBM-IBM method introduced in section~\ref{subsec:numerical_validation}.
Since there is no flux through the side pores, we can regard the SC pore as a one-in-one-out pore and compare to the reference distribution (\ref{eq:singlePoreConstantApproximation}) from section~\ref{subsec:varyingOneInOneOut}.
The results are presented in Fig. \ref{fig:SC_statistics_prediction}.

\begin{figure}[htbp]
  \centering
  \includegraphics[page=1,width=0.98\linewidth,trim=6 170 15 15,clip]{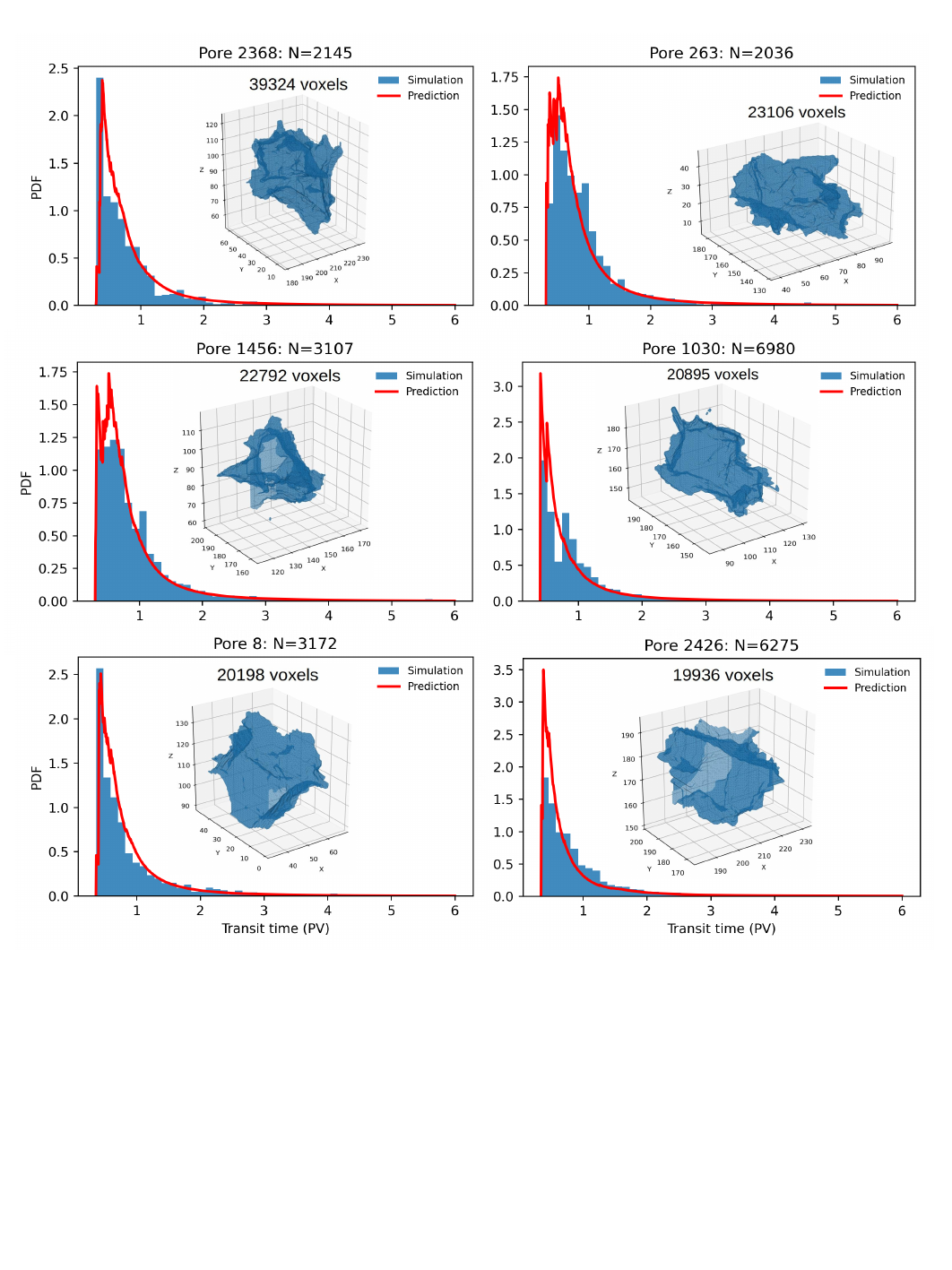}
  \caption{Transit time distribution in pores.
  Histograms are statistical results from simulation of one-batch injection and red curves are minD-PNM prediction. 
  Results are shown for large pores (volume $>18000$ voxels) that were traversed by plenty of particles (number $N>2000$) out of $50000$ in total. 
  Red lines are the predictions from equation~(\ref{eq:pdf_pore}).
  Other transit time distributions are shown in Appendix~\ref{app:pore-scale_TTD_figures}.}
  \label{fig:Quartz_Porous_Distribution}
\end{figure}

To assess the accuracy of the model prediction in a real medium, we conduct LBM-IBM simulations as introduced in section~\ref{subsec:numerical_validation} to simulate particle-laden flow transport through the porous medium shown in Fig.~\ref{fig:geometryPorousMedia}.
These simulations allow us to record the transit time distribution of particles in the total medium, but also in each individual pore.
Fig.~\ref{fig:Quartz_Porous_Distribution} shows that the pore-scale transit time distribution and the corresponding simulation results are in good agreement for several large pores. 
Thus, the subdomain division algorithm can accurately capture the pore-scale transit time distribution confirming the validity of Eq.~(\ref{eq:pdf_pore}).

\subsection{Predicting medium-scale transit time distribution and breakthrough curve for small samples}

\begin{figure}[H]
    \centering
    \begin{subfigure}[b]{0.44\linewidth}
        \centering
        \includegraphics[width=\linewidth]{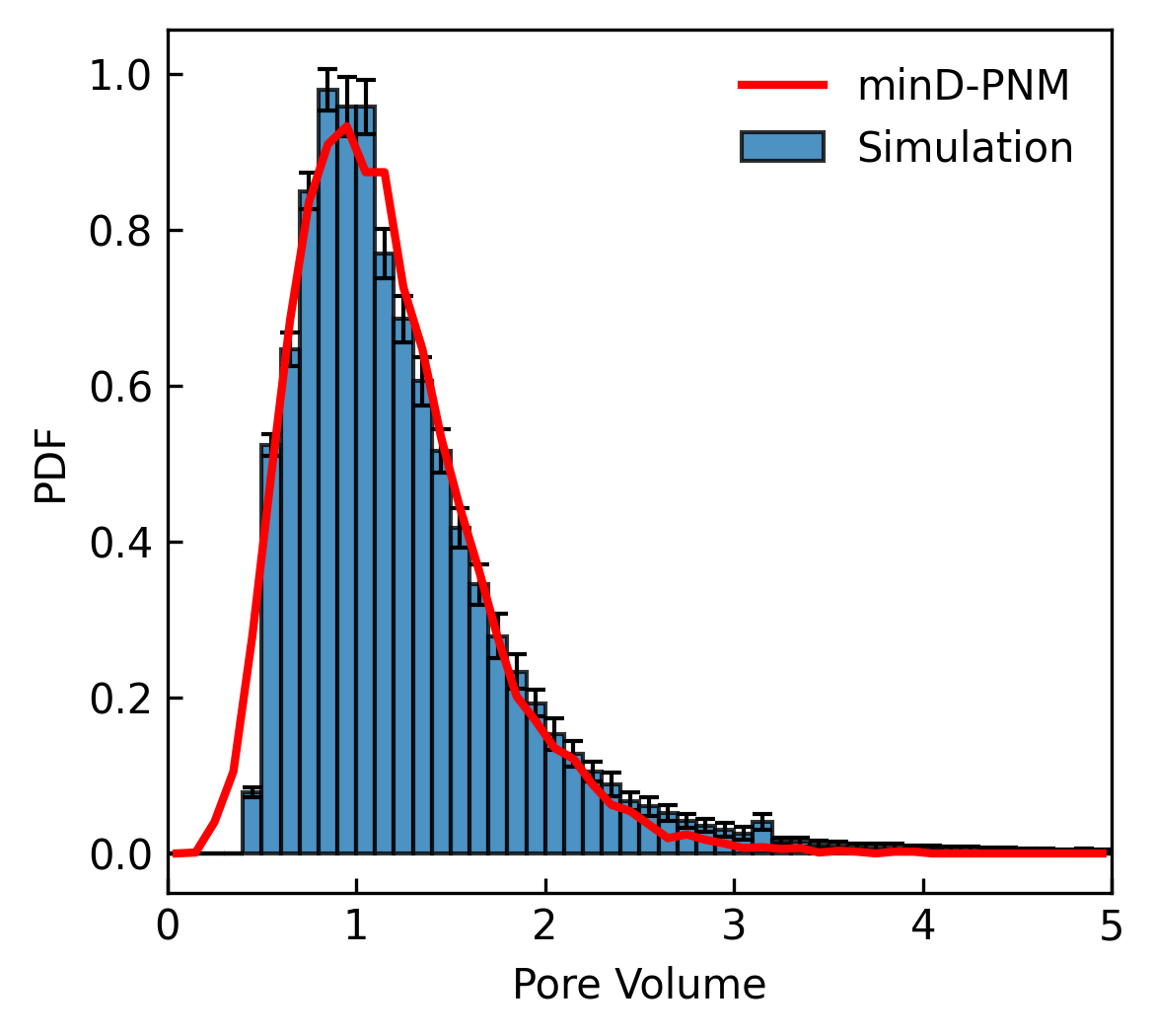}  \caption{}        \label{fig:pdf_PNM_oneLayer_hist}
    \end{subfigure}
    \hfill
    \begin{subfigure}[b]{0.54\linewidth}
        \centering
        \includegraphics[width=\linewidth]{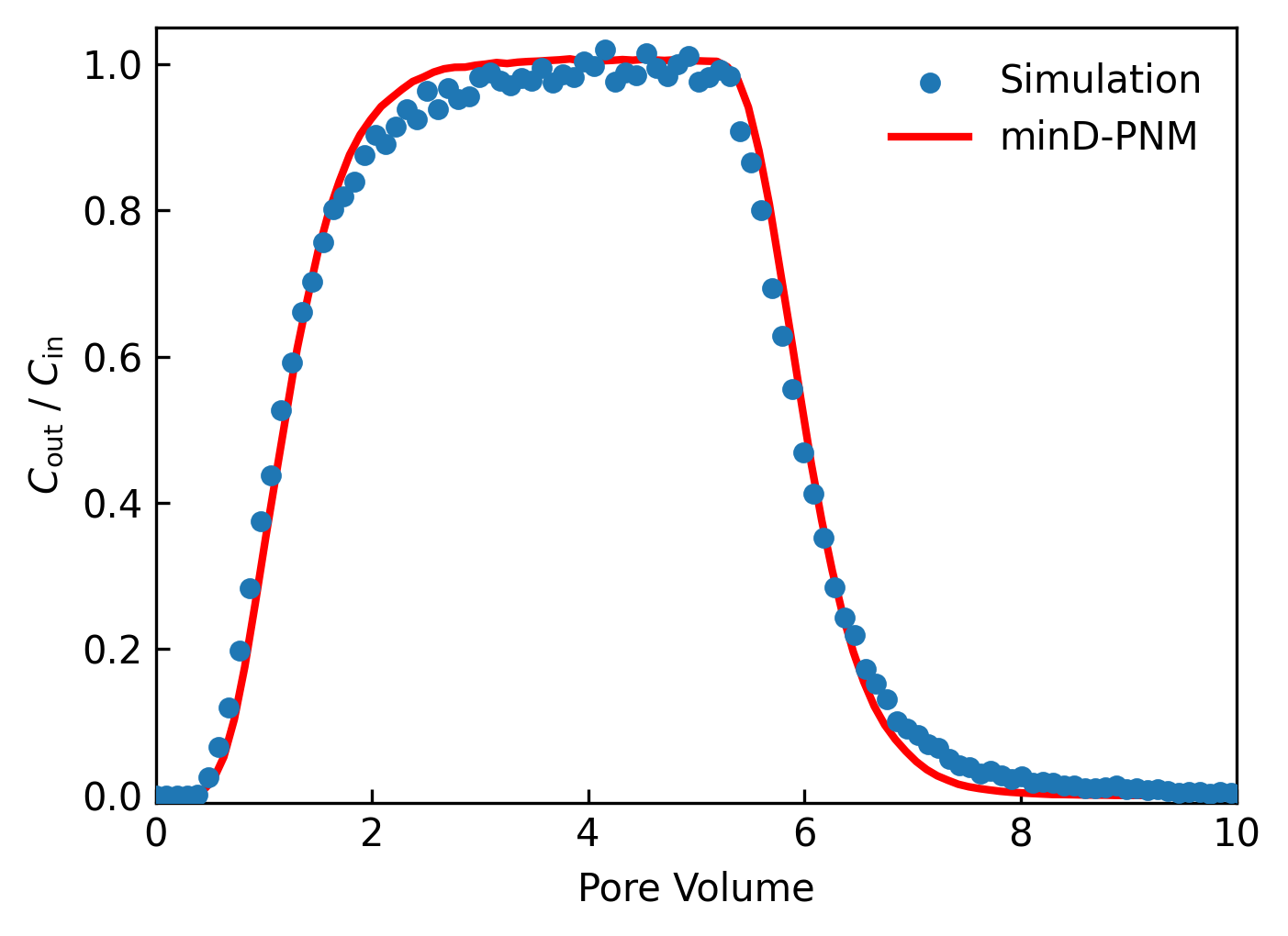}
        \caption{}
        \label{fig:BTCs}
    \end{subfigure}
    \caption{\centering Comparison between minD-PNM predictions and direct particle-resolved simulations:
    (a) medium-scale transit time distribution. The simulation results represent the average over 50 independent single-batch injections with randomly initialized particle positions, and the error bars indicate the standard error of the mean; (b) breakthrough curves obtained from particle-resolved simulations with 50 sequentially injected particle batches and from the minD-PNM model.}
    \label{fig:combined_BTC_PNM}
\end{figure}

Moving from a single pore to a full medium, Fig.~\ref{fig:combined_BTC_PNM} compares the transit time distributions and breakthrough curves predicted by the minD-PNM with those obtained from direct particle-based simulations for microplastic transport through the small sample shown in Fig.\ref{fig:geometryPorousMedia}. 
Table~\ref{tab:errors_PNM_models} quantifies the agreement using the coefficient of determination ($R^2$) and the root mean square error (RMSE). The calculation of these error metrics is described in Appendix~\ref{app:def_errors}.
We conclude from these observations that the minD-PNM predictions are in very good agreement with the direct numerical simulations.

\begin{table}[t]
    \centering
    \caption{Prediction errors of the minD-PNM model relative to direct simulations of the TTD and BTC.}    \label{tab:errors_PNM_models}
    \setlength{\tabcolsep}{14pt}    \renewcommand{\arraystretch}{1.2}
    \begin{tabular}{|c|
        S[table-format=1.4]
        S[table-format=1.4]|}
        \hline
        \textbf{Curve} & \multicolumn{1}{c}{\boldmath$R^2$}
        &\multicolumn{1}{c|}{\textbf{RMSE}} \\
        \hline
        TTD & 0.9795 & 0.0428 \\
        BTC & 0.9945 & 0.0316 \\
        \hline
    \end{tabular}
\end{table}

Having assessed the accuracy of our model, we move on to compare the computational cost.
Tab.~\ref{tab:comp_costs} contains the computational costs of different models for the same cubic quartz sample used in Fig.~\ref{fig:geometryPorousMedia}. 
The direct simulation (DS) relies on cluster computing and  provides full resolution of particle motion, flow field, and pore geometry, but requires the highest computational resources, with an estimated peak GPU memory requirement of approximately 1.05~GB and a runtime of 18~hours. 
In contrast, the pore-network model captures only the geometric connectivity but achieves extremely high efficiency, with a runtime of about 0.1~hours and negligible memory demand. 
These comparisons highlight the substantial computational advantage of the minD-PNM method for studying large-scale transport processes.

\begin{table}[htbp]
    \centering
    \caption{Comparison of the computational costs of direct simulation (DS) and minD-PNM for a cubic quartz sample with size $200 \times 6.5\,\mu\mathrm{m}$.}
    \label{tab:comp_costs} 

    \setlength{\tabcolsep}{4pt}
    \renewcommand{\arraystretch}{1.2}

    \begin{tabular}{|c|cccc|}
        \hline
        \textbf{Method}
        & \textbf{Particles/$10^3$}
        & \textbf{GPU-Memory/GB}
        & \textbf{Runtime/h}
        & \textbf{Resolved features} \\
        \hline
        DS       & 250 & 1.05 & 18  & G-P-F \\
        minD-PNM & 0  & 0     & 0.1 & N       \\
        \hline
    \end{tabular}

    \vspace{0.5em}

    \begin{minipage}{0.8\linewidth}
        \footnotesize
        G: geometry; P: particles; F: flow field; N: network.
    \end{minipage}
\end{table}

\subsection{Predicting breakthrough curves for macroscopic samples}
To assess the performance of our model in practice, we conducted microplastic breakthrough experiments using quartz sediment columns as introduced in section~\ref{subsec:experimental_validation}.
In this sample, local porosity varies with depth, and pores adjacent to container walls are typically larger due to looser packing, which makes it inaccurate to predict the transit time distribution using only a small subsample to represent the full geometry. 
Natural field sediments are similar but can be even more complex: pore size and connectivity vary with grain size, shape, and depth, while mixture materials and irregular boundaries further complicate the pore structure. 
Consequently, resolving transport over a representative column-scale pore domain becomes important, however, direct particle-resolved simulations over such domains can be computationally prohibitive.

The computational efficiency of minD-PNM allows us to perform such a column-scale prediction.
We use the $\mu$CT-resolved geometry of a quartz column, shown in Fig.~\ref{fig:geometryPorousMedia}~(a).
This $\mu$CT column is not the same physical column used in the breakthrough experiments, but it was prepared using the same type of quartz sediment with a comparable grain-size range.
It therefore provides an independent geometric packing realization representative of the quartz sediment system considered here. 

\begin{figure}[H]
    \centering
    \includegraphics[width=0.58\linewidth]{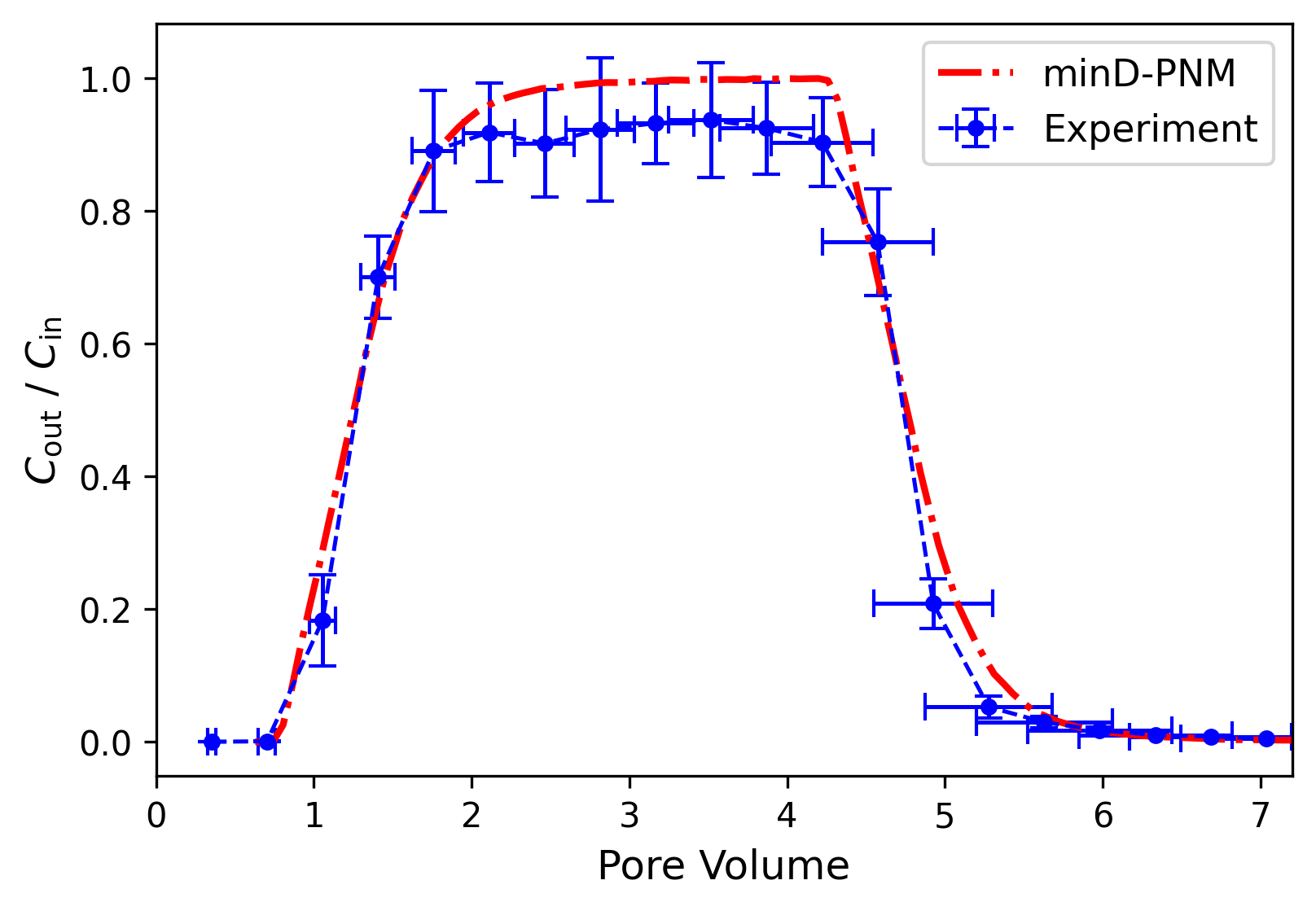}
    \caption{\centering Breakthrough curve of microplastic transport through quartz sediment columns compared with the minD-PNM prediction. Experimental values show the mean of six independent column experiments. Vertical error bars indicate the standard deviation of the normalized outlet concentration, while horizontal error bars indicate the standard deviation of the corresponding injected pore volumes.}
    \label{fig:BTC_PNM_Exp}
\end{figure}

We apply minD-PNM to this geometric realization to predict the breakthrough curve and compare the resulting response with the experimental measurements.
As shown in Fig.~\ref{fig:BTC_PNM_Exp}, the predicted BTC captures the characteristic rise and washout behavior observed experimentally. A systematic difference remains in the plateau concentration. However, this discrepancy can be expected because the present minD-PNM describes advection-dominated transport and does not include particle retention caused by particle-surface adsorption and straining resulting from particle aggregation, which exists in the experiment although dispersants were added to mitigate these effects. These processes reduce the fraction of particles reaching the outlet. The comparison therefore indicates that the geometry-driven model captures the principal advective spreading of microplastics through the quartz sediment.

\section{Conclusion}
We introduce a fast and memory-efficient simulation framework, minD-PNM, for predicting advection-dominated particle transport in complex, non-transparent porous media such as water-saturated quartz sediments. 
The method requires only the geometry of the porous medium as input and does not rely on any pre-computed fluid flow field. 
A key element of the method is the identification of an approximate decay law of $t^{-3}$ and the prediction of pore-scale transit time distributions by a subdomain division algorithm based on minimum energy dissipation. 
The method accurately characterizes transit times both in ideal sphere packings and in natural sediment geometries, which enables statistical modeling of sample-scale transport through efficient random-walk sampling and finally the prediction of breakthrough curves which are the central observable in laboratory experiments.
Compared to particle-resolved simulations, our method achieves substantial reductions in both computational cost and memory usage, lowering the barrier for large-scale modeling by several orders of magnitude.
We validate our method by comparing to direct numerical simulations including explicit particles for small samples and to experimental measurements on macroscopic quartz samples for which particle-resolved simulations are unfeasible.
With this we show that the minD-PNM method provides a scalable, physically grounded, and versatile approach for modeling particle dynamics in porous media.

\section{Acknowledgments}
This work was funded by the Deutsche Forschungsgemeinschaft (DFG, German Research Foundation) – SFB 1357 – Project number 391977956. Rizwan Khaleel was supported by the German Academic Exchange Service (DAAD) under the Graduate School Scholarship Program, 2023 (57650678). 
Calculations were performed using the festus-cluster of the Bayreuth Centre for High Performance Computing (https://www.bzhpc.uni-bayreuth.de), funded by the Deutsche Forschungsgemeinschaft (DFG, German Research Foundation) - 523317330.
Yifan Lu participated in the early development of the experimental method. The authors acknowledge the CRC 1357 Microplastics service project Z01, led by Prof. Dr. Seema Agarwal and Dr. Daniel Wagner, for providing the microplastic particles used in the experiments.

\newpage
\appendix
\section{Derivation of transit time distribution from velocity field\label{app:velocityToTTD}}

As defined in Section~\ref{subsec:TTD_oneinoneout}, flux-weighted injection distributes particles over the cross section according to the local volumetric flux, resulting in the probability density $\dfrac{\psi(\mathbf r)}{A}$. Since the corresponding dimensionless transit time is $t=\dfrac{1}{\psi(\mathbf r)}$, the probability density $\eta(t)$ of particle transit times can be written as
\begin{equation}
\eta(t)
=
\frac{1}{A}
\int_{\Omega}
\psi(\mathbf r)\,
\delta\!\left(
t-\frac{1}{\psi(\mathbf r)}
\right)
\,dA .
\label{eq:eta_delta_start}
\end{equation}

Using the co-area formula for Dirac function,
\begin{equation}
\int_{\Omega} f(\mathbf r)\,
\delta\!\left(h(\mathbf r)\right)\,dA
=
\int_{\{h=0\}}
\frac{f(\mathbf r)}
{|\nabla h(\mathbf r)|}\,d\ell,
\end{equation}
with $h(\mathbf r)=t-\dfrac{1}{\psi(\mathbf r)}$, the integral over the cross section can be transformed into an integral along the level set $\{\psi=\frac{1}{t}\}$, i.e. the contour of positions having the same velocity and therefore the same transit time:
\begin{equation}
\eta(t)
=
\frac{1}{A}
\int_{\{\psi=1/t\}}
\frac{\psi(\mathbf r)}
{\left|\nabla\!\left(\psi(\mathbf r)^{-1}\right)\right|}
\,d\ell .
\end{equation}
Since $\left|\nabla\!\left(\psi^{-1}\right)\right|=\dfrac{|\nabla\psi|}{\psi^2}$,
we obtain
\begin{equation}
\eta(t)
=
t^{-3}\,
\frac{1}{A}
\int_{\{\psi=1/t\}}
\frac{d\ell}{|\nabla\psi(\mathbf r)|},
\qquad
t\geq \frac{1}{\psi_{\max}} .
\label{eq:eta_levelset_final}
\end{equation}

Thus, the flux-weighted transformation from velocity to transit time produces the factor $t^{-3}$, while the remaining level-set integral accounts for the dependence on the cross-sectional velocity profile.

\section{Transit time distribution in elliptic Poiseuille flow}
\label{app:ellipticPoiseuille}

Consider an elliptic channel with cross section
\begin{equation}
\Omega=\left\{(x,y):\frac{x^2}{a^2}+\frac{y^2}{b^2}\leq1\right\}.
\end{equation}
The fully developed velocity shape function satisfying
$-\nabla^2\tau=1$ in $\Omega$ and $\tau=0$ on $\partial\Omega$ is
\begin{equation}
\tau(x,y)
=
\tau_{\max}
\left(
1-\frac{x^2}{a^2}-\frac{y^2}{b^2}
\right),
\qquad
\tau_{\max}
=
\frac{a^2b^2}{2(a^2+b^2)}.
\label{eq:tau_ellipse}
\end{equation}

Introducing elliptic coordinates
\[
x=ar\cos\theta,
\qquad
y=br\sin\theta,
\qquad
0\leq r\leq1,
\qquad
0\leq\theta<2\pi,
\]
with $dA=ab\,r\,dr\,d\theta$ and $A=\pi ab$, Eq.~\eqref{eq:tau_ellipse} becomes
\[
\tau(r)=\tau_{\max}(1-r^2).
\]
Its area average is
\begin{equation}
\langle\tau\rangle_A
=
\frac{1}{A}\int_{\Omega}\tau\,dA
=
\frac{\tau_{\max}}{2},
\end{equation}
and therefore the normalized velocity profile defined in the main text is
\begin{equation}
\psi(r)
=
\frac{\tau(r)}{\langle\tau\rangle_A}
=
2(1-r^2).
\label{eq:psi_ellipse}
\end{equation}
Importantly, this normalized profile is independent of the semi-axes $a$ and $b$, and therefore of the aspect ratio of the ellipse.

Using the flux-weighted transit time distribution derived in Appendix~\ref{app:velocityToTTD},
\begin{align}
\eta(t)
&=
\frac{1}{A}
\int_{\Omega}
\psi(r)
\delta\!\left(
t-\frac{1}{\psi(r)}
\right)dA
\nonumber\\
&=
4\int_0^1
(1-r^2)r\,
\delta\!\left(
t-\frac{1}{2(1-r^2)}
\right)dr .
\end{align}
With the substitution $u=1-r^2$, this becomes
\begin{equation}
\eta(t)=2\int_0^1 u\,\delta\!\left(t-\frac{1}{2u}
\right)du=\frac{1}{2}t^{-3},\quad t\geq\frac{1}{2}.
\label{eq:eta_ellipse}
\end{equation}

Thus, the flux-weighted transit time distribution of elliptic Poiseuille flow is exactly
\begin{equation}
\eta(t)=\frac{1}{2}t^{-3},\quad t\geq\frac{1}{2}
\end{equation}
independently of the elliptic aspect ratio.

\section{Numerical evaluation of transit time distributions for one-in-one-out pores}
\label{app:Numerical_method_PDF_singlePore}

To evaluate the transit time distribution for arbitrary cross-sectional geometries, we solve the Poisson problem introduced in Section~\ref{subsec:TTD_oneinoneout},
\begin{equation}
    -\nabla^2\tau = 1\;\; \mathrm{in} \;\; \Omega,
    \qquad
    \tau = 0\;\; \mathrm{on}\;\; \partial\Omega ,
    \label{eq:app_poisson_tau}
\end{equation}
on a uniform Cartesian grid with spacing $h$. Grid cells whose centers lie inside $\Omega$ are treated as fluid, and Eq.~\eqref{eq:app_poisson_tau} is discretized using the standard five-point stencil,
\begin{equation}
    4\tau_{m,n}-\tau_{m+1,n}-\tau_{m-1,n}-\tau_{m,n+1}-\tau_{m,n-1}=h^2 .
    \label{eq:app_fd_poisson}
\end{equation}
Values outside the pore domain are set to zero, corresponding to the no-slip boundary condition.

After solving the resulting sparse linear system, the velocity profile is normalized consistently with Eq.~\eqref{eq:defPsi} as
\begin{equation}
    \psi_{m,n}=
    \frac{\tau_{m,n}}
    {\langle\tau\rangle_h},
    \qquad
    \langle\tau\rangle_h
    =
    \frac{1}{N_f}
    \sum_{(m,n)\in\Omega_h}\tau_{m,n},
    \label{eq:app_psi_normalization}
\end{equation}
where $\Omega_h$ denotes the set of fluid cells and $N_f$ their number. The corresponding dimensionless transit time is $t_{m,n}=\dfrac{1}{\psi_{m,n}}$.

Under flux-weighted injection, each grid cell contributes with probability weight $w_{m,n}=\dfrac{\psi_{m,n}}{N_f}$, which satisfies $\sum\limits_{(m,n)\in\Omega_h}w_{m,n}=1$.
The numerical transit time distribution is then obtained from a weighted histogram. For a time bin $[t_k,t_{k+1})$,
\begin{equation}
    \eta_h(t_k)
    =
    \frac{1}{\Delta t_k}
    \sum_{\substack{(m,n)\in\Omega_h\\
    t_k\leq t_{m,n}<t_{k+1}}}
    w_{m,n},
    \qquad
    \Delta t_k=t_{k+1}-t_k .
    \label{eq:app_histogram_pdf}
\end{equation}
This is the discrete counterpart of the flux-weighted transit time distribution derived in Appendix~\ref{app:velocityToTTD} and avoids explicit extraction of the level sets $\left\{\psi=\dfrac{1}{t}\right\}$.

The numerical far tail requires particular care. Because $\psi \to 0$ at a no-slip boundary, $t = \dfrac{1}{\psi}$ becomes increasingly sensitive to discretization errors close to the wall. This sensitivity is further amplified by the stair-step representation of curved boundaries on the Cartesian grid. To avoid displaying this resolution-sensitive extreme tail, we restrict the histogram range to transit times below the 95th percentile of the values $t_{m,n} = \dfrac{1}{\psi_{m,n}}$ evaluated over all fluid grid cells. The percentile is defined with respect to the grid cells rather than the flux-weighted probability. The histogram weights remain normalized by the total flux weight over the complete cross section, such that no renormalization is applied after restricting the displayed transit-time range. This procedure removes only the numerically least reliable extreme tail from the comparison while retaining the resolved part of the transit time distribution shown in Fig.~\ref{fig:TTD_intermedia_region}.

The same discretization, flux weighting, and truncation procedure is applied to all cross-sectional geometries considered in the main text, including regular polygons, rectangles with different aspect ratios, and sinusoidally perturbed circular boundaries.

\section{Error analysis for the extension from straight to varying pores}
\label{App:ErrAnalysisSs}

\subsection{Eulerian approximation of the Lagrangian transit time}

For a slowly varying one-in-one-out pore, the exact transit time is a Lagrangian quantity obtained by following a streamline through the pore.
Let $\alpha\in[0,1]$ label streamlines according to the flux-weighted injection, and let $\psi_\alpha(s)$ denote the normalized local axial velocity along streamline $\alpha$ at position $s$.

Using the leading-order velocity
\[u_0(s,\mathbf r)=\frac{Q_\mathrm{p}}{A(s)}\psi(s,\mathbf r),\]
the dimensionless transit time along streamline $\alpha$, normalized by the pore-volume time $\dfrac{V_\mathrm{p}}{Q_\mathrm{p}}$, is
\begin{equation}
t_\alpha =\frac{1}{V_\mathrm{p}}\int_0^L \frac{A(s)}{\psi_\alpha(s)} \,ds .
\label{eq:t_alpha_appendix}
\end{equation}
For a straight pore, $\psi_\alpha$ is independent of $s$ and Eq.~\eqref{eq:t_alpha_appendix} reduces directly to $t_\alpha=\dfrac{1}{\psi_\alpha}$.

In a slowly varying pore, the normalized velocity profile changes only weakly along the axial direction. We therefore write
\begin{equation}
\psi_\alpha(s)=\bar{\psi}_\alpha\left[1+\mathcal O(\varepsilon_s)\right],
\label{eq:psi_alpha_variation_appendix}
\end{equation}
where $\varepsilon_s\ll1$ characterizes the axial variation of the normalized velocity profile. Substitution into Eq.~\eqref{eq:t_alpha_appendix}, together with $V_\mathrm{p}=\displaystyle\int_0^L A(s)\,ds$, gives
\begin{equation}
t_\alpha=\frac{1}{\bar{\psi}_\alpha}\left[1+\mathcal O(\varepsilon_s)\right].
\label{eq:t_alpha_error_appendix}
\end{equation}

The Eulerian approximation used in the main text therefore differs from the exact streamline-based transit time only by a relative correction of order $\mathcal O(\varepsilon_s)$ when the cross-sectional velocity profile varies weakly along the pore. Since the flux-weighted transit time distribution can be written as
\begin{equation}
\eta(t)=\int_0^1 \delta\!\left(t-t_\alpha\right) \,d\alpha ,
\label{eq:eta_Lag_appendix}
\end{equation}
this correction perturbs the transit time distribution only at higher order and does not alter its leading $t^{-3}$ scaling.

\subsection{Higher-order correction from the lubrication approximation}

We next estimate the error introduced by replacing the exact Stokes flow with the local Poiseuille approximation. As defined in the main text, we consider a slender and slowly varying pore with
\[\varepsilon\sim\frac{R_\mathrm{t}}{L}\ll1,\]
where $R_\mathrm{t}$ is the characteristic transverse pore size and $L$ is the pore length. The slow cross-sectional variation is characterized by a representative wall slope satisfying $\left|\dfrac{dR_\mathrm{t}}{ds}\right| = O(\varepsilon)$, while the weak centerline curvature is characterized by $\kappa R_{\mathrm{t}} = O(\varepsilon)$.
The axial derivatives scale as $\partial_s\sim\dfrac{\varepsilon}{R_\mathrm{t}}$, whereas transverse derivatives scale as $\nabla_{\mathbf r}\sim\dfrac{1}{R_\mathrm{t}}$. Hence the axial viscous term is smaller than the transverse viscous term by a factor of
$\mathcal O(\varepsilon^2)$.

The leading-order axial velocity therefore satisfies the local cross-sectional Poisson problem,
\begin{equation}
-\frac{dp_0}{ds}+\mu\nabla_{\mathbf r}^2u_0=0
\;\;\mathrm{in}\;\; \Omega(s),
\quad
u_0=0\;\;\mathrm{on}\;\; \partial\Omega(s).
\label{eq:leading_poisson_appendix}
\end{equation}
Since the $O(\varepsilon)$ correction has the same cross-sectional form as the leading-order solution, while the prescribed flux $Q$ is already carried by $u_0$, flux conservation requires the first-order correction to vanish. Therefore, the full axial velocity differs from this local solution by
\begin{equation}
u_s=u_0\left[1+\mathcal O(\varepsilon^2)\right].
\label{eq:u_u0_eps2_appendix}
\end{equation}

For a weakly curved pore, the streamline also differs slightly from the pore centerline. If its transverse displacement satisfies $\left|\dfrac{d\mathbf r_b}{ds}\right|=\mathcal O(\varepsilon)$,
then its arc-length element is
\begin{equation}
ds_\alpha=\sqrt{1+\left|\frac{d\mathbf r_b}{ds}\right|^2
}\,ds
=\left[1+\mathcal O(\varepsilon^2)\right]ds .
\label{eq:ds_alpha_appendix}
\end{equation}
Combining this result with Eq.~\eqref{eq:u_u0_eps2_appendix} gives
\begin{equation}
\int_0^L\frac{ds_\alpha}{u_s}=
\int_0^L\frac{ds}{u_0}\left[1+\mathcal O(\varepsilon^2)\right].
\label{eq:transit_time_eps2_appendix}
\end{equation}
Thus, under the slowly varying and weak-curvature assumptions adopted here, the lubrication approximation introduces only a relative $\mathcal O(\varepsilon^2)$ correction to the streamline transit time.

In summary, extending the straight-pore result to a slowly varying one-in-one-out pore introduces two subleading corrections: an $\mathcal O(\varepsilon_s)$ error associated with replacing the exact Lagrangian evolution of the normalized velocity by its Eulerian approximation, and an $\mathcal O(\varepsilon^2)$ error associated with the local Poiseuille approximation and streamline geometry. Neither
correction changes the leading $t^{-3}$ scaling in the slowly varying limit.

\section{Error metrics for model validation}
\label{app:def_errors}

The agreement between the minD-PNM predictions and the corresponding reference results is quantified using the coefficient of determination ($R^2$) and the root mean square error (RMSE). The metrics are evaluated directly from the quantities shown in the transit-time distribution (TTD) and breakthrough curve (BTC) comparisons.

For the TTD comparison, the probability density predicted by minD-PNM, $\eta_{\mathrm{PNM}}(t_k)$, is compared with that obtained from the particle-resolved simulation, $\eta_{\mathrm{sim}}(t_k)$, in the same transit-time bin $t_k$. The coefficient of determination is
\begin{equation}
R^2_{\mathrm{TTD}}=
1-\frac{\displaystyle\sum_{k=1}^{N_{\mathrm{TTD}}}\left[\eta_{\mathrm{sim}}(t_k)-\eta_{\mathrm{PNM}}(t_k)\right]^2}
{\displaystyle\sum_{k=1}^{N_{\mathrm{TTD}}}\left[\eta_{\mathrm{sim}}(t_k)-\overline{\eta}_{\mathrm{sim}}\right]^2},
\end{equation}
where
\begin{equation}
\overline{\eta}_{\mathrm{sim}}=
\frac{1}{N_{\mathrm{TTD}}}\sum_{k=1}^{N_{\mathrm{TTD}}}\eta_{\mathrm{sim}}(t_k),
\end{equation}
and $N_{\mathrm{TTD}}$ is the number of transit-time bins included in the comparison. The corresponding RMSE is
\begin{equation}
\mathrm{RMSE}_{\mathrm{TTD}}=
\sqrt{\frac{1}{N_{\mathrm{TTD}}}\sum_{k=1}^{N_{\mathrm{TTD}}}\left[\eta_{\mathrm{sim}}(t_k)-\eta_{\mathrm{PNM}}(t_k)\right]^2}.
\end{equation}

For the BTC comparison, we denote the normalized reference and minD-PNM curves by $C_{r,\mathrm{ref}}(t)$ and $C_{r,\mathrm{PNM}}(t)$, respectively. Here, ``ref'' denotes either the particle-resolved simulation or the experiment, depending on the comparison. For the experimental data, the relative concentration is defined as the measured outlet concentration divided by the corresponding inlet concentration. For the particle-resolved simulation, it is defined as the number of particles reaching the outlet during each sampling interval normalized by the number of particles injected per batch.

The coefficient of determination for the BTC is
\begin{equation}
R^2_{\mathrm{BTC}}=
1-\frac{\displaystyle\sum_{k=1}^{N_{\mathrm{BTC}}}\left[C_{r,\mathrm{ref}}(t_k)-C_{r,\mathrm{PNM}}(t_k)
\right]^2}
{\displaystyle\sum_{k=1}^{N_{\mathrm{BTC}}}\left[C_{r,\mathrm{ref}}(t_k)-\overline{C}_{r,\mathrm{ref}}\right]^2},
\end{equation}
where
\begin{equation}
\overline{C}_{r,\mathrm{ref}}=
\frac{1}{N_{\mathrm{BTC}}}\sum_{k=1}^{N_{\mathrm{BTC}}}C_{r,\mathrm{ref}}(t_k),
\end{equation}
and $N_{\mathrm{BTC}}$ is the number of sampling points included in the comparison. The corresponding RMSE is
\begin{equation}
\mathrm{RMSE}_{\mathrm{BTC}}=
\sqrt{\frac{1}{N_{\mathrm{BTC}}}\sum_{k=1}^{N_{\mathrm{BTC}}}\left[C_{r,\mathrm{ref}}(t_k)-C_{r,\mathrm{PNM}}(t_k)\right]^2}.
\end{equation}

In both cases, the prediction and reference values are compared at identical transit-time bins or sampling times. The $R^2$ value quantifies the fraction of the variation in the reference data reproduced by the model, whereas the RMSE represents the characteristic magnitude of the pointwise prediction error.

\section{Pore-scale transit time distribution in pores with more than $15000$ voxels\label{app:pore-scale_TTD_figures}}

\begin{figure}[htbp]
  \centering
  \includegraphics[page=1,width=0.95\linewidth,trim=10 4 10 10,clip]{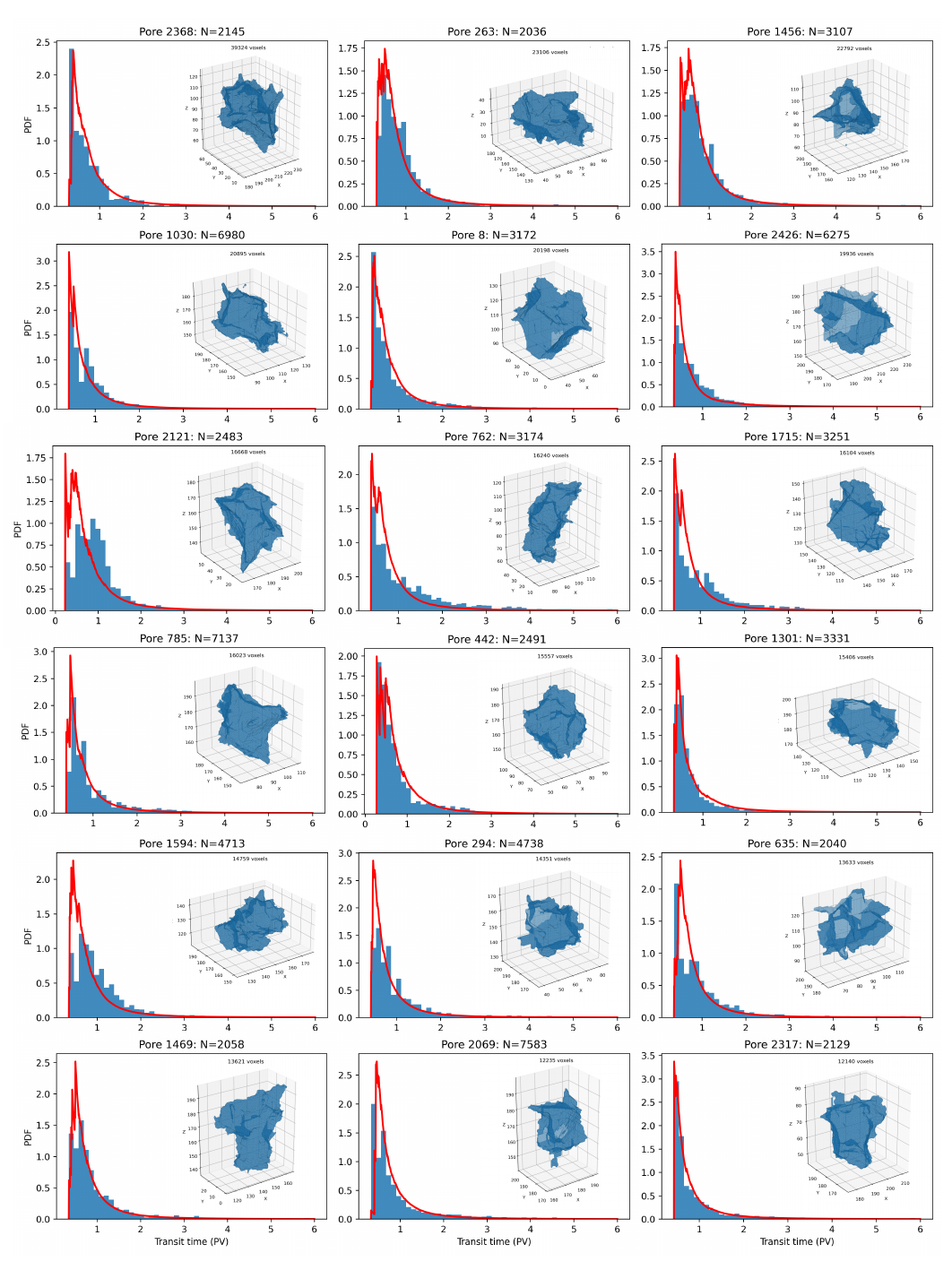}
  \label{fig:Quartz_Porous_Distribution1}
  \caption{Additional transit time distributions in individual pores.}
\end{figure}

\pagebreak

\section{Additional methods for the microplastic transport experiment\label{app:transport_experiment}}

\begin{table}[H]
\centering
\caption{Composition of artificial rainwater used in the transport experiments.}
\label{tab:artificial_rainwater}
\begin{tabular}{lc}
\toprule
\textbf{Salt} & \textbf{Concentration (mg\,L$^{-1}$)} \\
\midrule
(NH$_4$)$_2$SO$_4$ & 0.488 \\
CaSO$_4$           & 0.234 \\
NH$_4$NO$_3$       & 0.654 \\
CaCl$_2$           & 0.154 \\
Na$_2$CO$_3$       & 0.244 \\
\bottomrule
\end{tabular}
\end{table}

\begin{figure}[H]
  \centering
  \includegraphics[page=1,width=\textwidth]{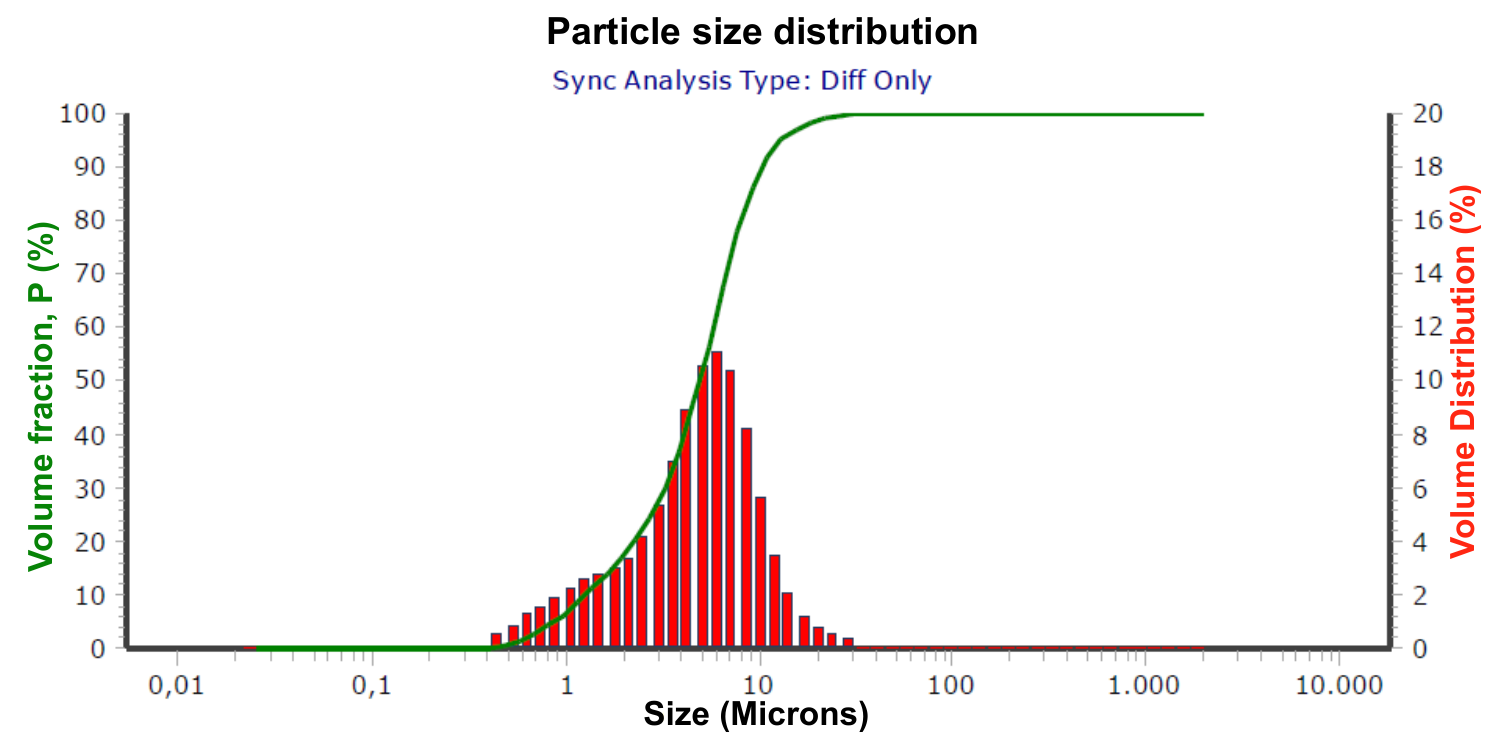}
  \caption{Particle-size distribution of Rhodamine B-stained polystyrene (RhB-PS) microplastic particles ($D_{90}=10~\mu$m) measured using a Microtrac FlowSync system. The RhB-PS particles were produced in-house by CRC 1357 Project Z01 (University of Bayreuth) from polystyrene raw material (Magic Pyramid GmbH, Germany). Briefly, a 2 wt\% polystyrene solution in tetrahydrofuran (THF) was spray-dried at SKZ Selb (S\"uddeutsches Kunststoff-Zentrum, Selb, Germany), resulting in RhB-PS particles with a characteristic particle-size distribution of $D_{90}=10~\mu$m.}
  \label{fig:Particle_Size_Distribution}
\end{figure}

\newpage
\section*{references}
\bibliography{apssamp}

\end{document}